\documentclass[final,authoryear,3p,times]{elsarticle}

\usepackage{booktabs}
\usepackage{amsmath,amssymb}
\usepackage{graphicx}
\usepackage[colorlinks]{hyperref}
\usepackage{natbib}

\newcommand{\boldface}[1]{\boldsymbol{#1}}
\newcommand{\bfa}{\boldface{a}}

\newcommand{\bfe}{\boldface{e}}

\newcommand{\bfn}{\boldface{n}}

\newcommand{\bfp}{\boldface{p}}

\newcommand{\bft}{\boldface{t}}
\newcommand{\bfu}{\boldface{u}}
\newcommand{\bfv}{\boldface{v}}

\newcommand{\bfx}{\boldface{x}}

\newcommand{\bfI}{\boldface{I}}
\newcommand{\bfT}{\boldface{T}}
\newcommand{\bfepsilon}{\boldsymbol{\varepsilon}}
\newcommand{\bfsigma}{\boldsymbol{\sigma}}

\newcommand{\bfnull}{\boldsymbol{0}}
\newcommand{\bfGamma}{\boldsymbol{\Gamma}}

\newcommand{\dsC}{\mathbb{C}}
\newcommand{\dsS}{\mathbb{S}}
\newcommand{\T}{^{\mathsf{T}}}
\newcommand{\Rset}{\ensuremath{\mathbb{R}}}
\def\dd{\;\!\mathrm{d}}
\def\id{{\bfI}}
\DeclareMathOperator{\divv}{div}
\DeclareMathOperator{\tr}{tr}
\DeclareMathOperator{\grad}{grad}
\DeclareMathOperator{\argmin}{{arg\,min}}
\DeclareMathOperator{\sym}{sym}
\newcommand{\be}{\begin{equation}}
\newcommand{\ee}{\end{equation}}

\begin{document}

\begin{frontmatter}

\title{Rigorous bounds on the effective moduli of composites and inhomogeneous bodies with negative-stiffness phases}

\author[dmk]{Dennis M. Kochmann\corref{cor1}}
\ead{kochmann@caltech.edu}
\cortext[cor1]{Corresponding author (phone +1-626-395-8113, fax +1-626-395-2900).}
\author[gwm]{Graeme W. Milton}
\ead{milton@math.utah.edu}
\address[dmk]{Graduate Aerospace Laboratories, California Institute of Technology, Pasadena, CA 91125, USA}
\address[gwm]{Department of Mathematics, The University of Utah, Salt Lake City, UT 84112, USA}

\begin{abstract}
We review the theoretical bounds on the effective properties of linear elastic inhomogeneous solids (including composite materials) in the presence of constituents having non-positive-definite elastic moduli (so-called negative-stiffness phases). We show that for statically stable bodies the classical displacement-based variational principles for Dirichlet and Neumann boundary problems hold but that the dual variational principle for traction boundary problems does not apply. We illustrate our findings by the example of a coated spherical inclusion whose stability conditions are obtained from the variational principles. We further show that the classical Voigt upper bound on the linear elastic moduli in multi-phase inhomogeneous bodies and composites applies and that it imposes a stability condition: overall stability requires that the effective moduli do not surpass the Voigt upper bound. This particularly implies that, while the geometric constraints among constituents in a composite can stabilize negative-stiffness phases, the stabilization is insufficient to allow for extreme overall static elastic moduli (exceeding those of the constituents). Stronger bounds on the effective elastic moduli of isotropic composites can be obtained from the Hashin-Shtrikman variational inequalities, which are also shown to hold in the presence of negative stiffness.
\end{abstract}

\begin{keyword}
Stability \sep Elasticity \sep Composite \sep Negative Stiffness \sep Effective Properties \sep Bounds
\end{keyword}

\end{frontmatter}

\vspace{0.4cm}

\section{Introduction}

The overall or effective properties of heterogeneous solids are uniquely linked to the properties of each composite constituent, their geometric arrangement and bonding. Owing to microstructural randomness in the arrangement of composite phases, effective physical properties in most cases cannot be determined exactly. One approach is to estimate them by the aid of rigorous upper and lower bounds. The simplest such bounds were introduced by \citet{Hill1952} and \citet{Paul1966}: the Reuss and Voigt bounds are solely based on phase volume fractions and present upper and lower bounds on the effective linear elastic moduli of multi-phase composites. For inhomogeneous bodies the analogous bounds were obtained by~\citet{NematNasserHori1993} and by Willis in a 1989 private communication to Nemat-Nasser and Hori. Based on variational principles and the introduction of a polarization field, \citet{HashinShtrikman1963} derived new tighter bounds for isotropic well-ordered two-phase composites with bulk moduli $\kappa_2>\kappa_1$ and shear moduli $\mu_2>\mu_1$. Their bounds on the effective bulk modulus can be attained e.g.\ by assemblages of coated spheres (interchanging the materials in spherical inclusions and coatings yields upper and lower bounds on the effective bulk modulus). Similarly, hierarchical laminate constructions have been shown to attain the bounds on the effective shear modulus~\citep{Norris1985,Milton1986,Francfort1986}. Therefore, the bounds of \citet{HashinShtrikman1963} are optimal and present the strongest possible restrictions on the elastic moduli of well-ordered multi-phase solids based only on volume fractions. For non-well-ordered isotropic two-phase composites (with bulk moduli $\kappa_2>\kappa_1$ and shear moduli $\mu_2<\mu_1$) the tightest known bounds on the effective bulk modulus are those of \citet{Hill1963}, and are attained by coated-sphere assemblages, and on the effective shear modulus are those of~\citet{MiltonPhanThien1982}, which improve upon those of~\citet{Walpole1966} and are attained in certain parameter regimes where they coincide with the Hashin-Shtrikman formulae. By including statistical microstructural information of random composites, three-point bounds were derived e.g.\ by~\citet{BeranMolyneux1966} and \citet{McCoy1970}, who used classical variational principles. For two-phase composites these bounds were simplified by~\citet{Milton1981a}. By improving \mbox{McCoy's} bounds, \mbox{\citet{MiltonPhanThien1982}} found stronger restrictions for the effective shear modulus. We refer to \citep{Cherkaev2000,Torquato2002,Allaire2002,Milton2002,Tartar2010} for comprehensive reviews of composite bounds. Alternatively, estimates of effective composite properties have been established by an effective medium strategy, which has resulted in, among others, the self-consistent method~\citep{Hill1965,Budiansky1965,Berryman1980} and its generalized form~\citep{ChristensenLo1979}, the differential~\citep{Roscoe1952,Roscoe1973,Norris1985} and Mori-Tanaka schemes~\citep{MoriTanaka1973,Benveniste1987}. Although beyond the scope of the present investigation, we note that nonlinear variational bounds on composite properties are available as well, see e.g.~\citep{TalbotWillis1985,Castaneda1991,PonteCastanedaWillis1999}.

All of the aforementioned bounds imply that the effective linear elastic moduli of composites (in particular the Young, bulk, and shear moduli of isotropic composites) are bounded from above by the individual moduli of the constituent materials; i.e.\ no composite can be stiffer than its stiffest constituent. This prohibits the creation of new composites with extreme properties (where by `\emph{extreme}' we refer to properties which exceed those of the constituents). However, the derivation of those bounds assume that all constituent materials possess positive-definite elastic moduli (for the specific case of isotropic solids, this is equivalent to requiring Young, bulk and shear moduli to be positive). \citet{LakesDrugan2002} showed that relaxing this assumption by allowing for non-positive-definite elastic moduli (so-called negative stiffness) in one of the phases in an inhomogeneous body may lead to extreme effective stiffness. Based on exact solutions for a coated-sphere two-phase solid, they showed that a two-phase inhomogeneous body can, in principle (within the validity of the elasticity model), attain unbounded effective bulk stiffness if the constituent moduli and volume fractions are appropriately tuned. Lakes and coworkers demonstrated that various other effective physical composite properties promise to reach extreme values when including a negative-stiffness phase~\citep{Lakes2001a,Lakes2001b,WangLakes2001,WangLakes2004,WangLakes2005a}. Experimentally, negative stiffness has been realized by constituents undergoing microscale instabilities such as phase transitions, see e.g.~\citep{LakesEtAl2001,JaglinskiEtAl2006,JaglinskiLakes2007,JaglinskiEtAl2007}. Similarly, on a structural level the negative-stiffness effect has been realized by buckling instabilities, see e.g.~\citep{MooreEtAl2006,LeeEtAl2007,Lee2012,KashdanEtAl2012}.

While negative stiffness is generally unstable in homogeneous solids with mixed or pure-traction boundary conditions~\citep{Kirchhoff1859}, it was shown that negative-stiffness phases can be stabilized when geometrically constrained e.g.\ by a sufficiently stiff and thick coating or as inclusions in a stiff matrix~\citep{Drugan2007,KochmannDrugan2009,Kochmann2012,KochmannDrugan2012}. Unfortunately, the thus expanded stability regime is insufficient to stabilize extreme effective static stiffness in simple two-phase solids and in isotropic two-phase composites with equal shear moduli~\citep{WojnarKochmann2013a,WojnarKochmann2013b}, while allowing for interesting dynamic phenomena. 

Instead of investigating particular composite geometries, here we show that arbitrary linear elastic inhomogeneous bodies and multi-phase composites cannot reach extreme stiffness by the inclusion of negative-stiffness phases if they are to be statically stable. To this end, we first review the classical variational principles in Section~\ref{sec:VariationalPrinciples} and determine their validity in the presence of negative-stiffness phases. We illustrate the applicability or non-applicability of the various variational principles in Section~\ref{sec:CoatedSphere} by the example of a coated spherical inclusion. Next, in Section~\ref{sec:EffectiveModuli} we apply the variational principles to show that the classical Voigt upper bound applies and, most importantly, implies a stability condition: overall stability requires that the effective moduli must not surpass the Voigt upper bound. We further show that the Hashin-Shtrikman variational inequalities apply and that they yield additional upper and lower bounds on the effective elastic moduli of isotropic composites. Our results particularly demonstrate that extreme effective (static) elastic moduli exceeding those of any of the constituents are prohibited if the solid is to be statically stable. We further review three-point bounds on the effective moduli and conclude upper and lower bounds in the presence of negative-stiffness phases. Finally, Section~\ref{sec:Conclusions} concludes our analysis.


\section{Stability conditions for elastic solids}
\label{sec:VariationalPrinciples}

\subsection{Dirichlet problem: essential boundary conditions}
\label{sec:DirichletProblem}

We consider an inhomogeneous body $\Omega$ containing a composite material made of linear (visco)elastic constituents and experiencing a displacement field $\bfu(\bfx,t)$ with $\bfx\in\Rset^d$ denoting position in $d$-dimensional space and $t$ being time. Assume $\tilde{\bfu}(\bfx)$ is the displacement field corresponding to a solution of the elasticity equation (linear momentum balance) with essential boundary conditions $\tilde{\bfu}(\bfx)=\bfv(\bfx)$ on the body's boundary $\partial\Omega$ and given eigenstrains $\bfepsilon_0(\bfx)$ within the solid. The strain energy density of the linear elastic solid with locally-varying modulus tensor $\dsC(\bfx)$ is given by
\be
	\Psi(\bfepsilon) = \frac{1}{2}\left[\bfepsilon(\bfx)-\bfepsilon_0(\bfx)\right]\cdot \dsC(\bfx) \,
		\left[\bfepsilon(\bfx)-\bfepsilon_0(\bfx)\right],
\ee
where $\bfepsilon(\bfx) = \frac{1}{2}\left[\grad\bfu+(\grad\bfu)\T\right]$ is the infinitesimal symmetric strain tensor, $\bfepsilon_0(\bfx)$ is the infinitesimal symmetric eigenstrain tensor. We will see that a necessary condition for stability is that $\tilde{\bfu}(\bfx)$ minimizes the total elastic energy
\be \label{eq:infcondition}
	W = \inf_{\Large\bfu(\bfx) \atop \bfu(\bfx)=\bfv(\bfx) \ \text{on} \ \partial\Omega}
		\int_\Omega \Psi(\bfepsilon(\bfx)) \,\dd V.
\ee

To show this, assume that the solution $\tilde{\bfu}(\bfx)$ is not the minimizer of~\eqref{eq:infcondition} and that there is some $\hat{\bfu}(\bfx)$ which satisfies the essential boundary conditions and for which
\be\label{eq:assumption}
	\int_\Omega \Psi\left(\hat{\bfepsilon}(\bfx)\right) \dd V <
	\int_\Omega \Psi\left(\tilde{\bfepsilon}(\bfx)\right) \dd V.
\ee
Further, assume the inhomogeneous body has mass density $\rho(\bfx)>0$ and for simplicity is viscoelastic. Let the body have an initial displacement field at time $t=0$ given by
\be\label{eq:initialfield}
	\bfu(\bfx,0) = \bfu_\text{initial}(\bfx) = \frac{\tilde{\bfu}(\bfx) + \eta\,\hat{\bfu}(\bfx)}{1+\eta}
\ee
with some $\eta\in\Rset$. Note that~\eqref{eq:initialfield} satisfies the essential boundary conditions as well, i.e. $\bfu_\text{initial}(\bfx) = \bfv(\bfx)$ on $\partial\Omega$. We fix the displacements to be $\bfu_\text{initial}(\bfx)$ for all times $t\leq 0$ by application of appropriate body forces which we remove for all $t>0$, so that for $t\geq 0$ the body is out of equilibrium, i.e. 
\be\label{eq:displacements}
	\bfu(\bfx,t) = \begin{cases}
		\bfu_\text{initial}(\bfx) & \text{for} \ t\leq 0,\\
		\text{unknown} & \text{for} \ t> 0.
	\end{cases}	
\ee
We maintain essential boundary conditions
\be
	\bfu(\bfx,t) = \bfv(\bfx) \ \text{on} \ \partial\Omega
\ee
for all times $t$, so that no work is done on the body and either internal motions will be damped through viscosity and drive the solid into a state of stable equilibrium, or alternatively there will be no stable equilibrium. Displacements~\eqref{eq:displacements} result in strains $\bfepsilon_\text{initial}(\bfx)$ for $t\leq 0$, so that the initial energy of the body is purely elastic and equal to
\be
\begin{split}
	W_\text{initial} & = \int_\Omega \Psi\left(\bfepsilon_\text{initial}(\bfx)\right)\dd V\\
	 & = \int_\Omega \frac{1}{2}\,\frac{\tilde{\bfepsilon}(\bfx)-\bfepsilon_0(\bfx) 
	 	+ \eta\left[\hat{\bfepsilon}(\bfx)-\bfepsilon_0(\bfx)\right]}{1+\eta}\cdot\dsC(\bfx)\,
	 	\frac{\tilde{\bfepsilon}(\bfx)-\bfepsilon_0(\bfx) 
	 	+ \eta\left[\hat{\bfepsilon}(\bfx)-\bfepsilon_0(\bfx)\right]}{1+\eta}\,\dd V\\
	 & = \frac{1}{(1+\eta)^2} \left[
	 	\int_\Omega \Psi\left(\tilde{\bfepsilon}(\bfx)\right) \dd V
	 	+ \eta^2 \int_\Omega \Psi\left(\hat{\bfepsilon}(\bfx)\right) \dd V
	 	+ \eta\, \int_\Omega \left[\hat{\bfepsilon}(\bfx)-\bfepsilon_0(\bfx)\right]\cdot \dsC(\bfx)
	 		 \,\left[\tilde{\bfepsilon}(\bfx)-\bfepsilon_0(\bfx)\right] 
	 	\dd V\right].
\end{split}
\ee
Because $\tilde{\bfu}(\bfx)$ is a solution to the equilibrium equation, we use linear momentum balance in the absence of body forces and in static equilibrium, i.e. 
\be\label{eq:StatEqu}
	\divv\tilde{\bfsigma} = \divv \left[\dsC(\bfx)\,\left(\tilde{\bfepsilon}(\bfx)-\bfepsilon_0(\bfx)\right)\right]=\bfnull \quad \text{in} \ \Omega.
\ee
Utilizing symmetry of the infinitesimal stress tensor $\bfsigma$ and using~\eqref{eq:StatEqu}, we see that
\be
\begin{split}
	\int_\Omega \left[\hat{\bfepsilon}(\bfx)-\bfepsilon_0(\bfx)\right]\cdot \dsC(\bfx)
	 		 \,\left[\tilde{\bfepsilon}(\bfx)-\bfepsilon_0(\bfx)\right] 
	= & \int_\Omega \left[\hat{\bfepsilon}(\bfx)-\bfepsilon_0(\bfx)\right] \cdot \tilde{\bfsigma}(\bfx)\,\dd V
	= \int_\Omega \left[\grad\hat{\bfu}(\bfx)-\grad\bfu_0(\bfx)\right] \cdot \tilde{\bfsigma}(\bfx)\,\dd V \\
	= & \left[\int_{\partial\Omega} \left[\hat{\bfu}(\bfx)-\bfu_0(\bfx)\right] \cdot \tilde{\bfsigma}(\bfx)\bfn\,\dd S - 
		\int_\Omega \left[\hat{\bfu}(\bfx)-\bfu_0(\bfx)\right] \cdot \divv\tilde{\bfsigma}(\bfx)\,\dd V \right]\\
	= & \int_{\partial\Omega} \left[\hat{\bfu}(\bfx)-\bfu_0(\bfx)\right] \cdot \tilde{\bfsigma}(\bfx)\bfn\,\dd S
	= \int_{\partial\Omega} \left[\bfv(\bfx)-\bfu_0(\bfx)\right] \cdot \tilde{\bfsigma}(\bfx)\bfn\,\dd S \\ 
	= & \int_{\partial\Omega} \left[\tilde{\bfu}(\bfx)-\bfu_0(\bfx)\right] \cdot \tilde{\bfsigma}(\bfx)\bfn\,\dd S 
	= \int_\Omega \left[\tilde{\bfepsilon}(\bfx) - \bfepsilon_0(\bfx)\right] \cdot \tilde{\bfsigma}(\bfx)\,\dd V \\
	= & \int_\Omega \left[\tilde{\bfepsilon}(\bfx)-\bfepsilon_0(\bfx)\right]\cdot \dsC(\bfx)
	 		 \,\left[\tilde{\bfepsilon}(\bfx)-\bfepsilon_0(\bfx)\right]\,\dd V
	 	= 2 \int_\Omega \Psi(\tilde{\bfepsilon}(\bfx))\,\dd V.
\end{split}
\ee
Consequently,
\be
\begin{split}
	W_\text{initial} & = \frac{1}{(1+\eta)^2} \bigg[(1+2\eta) \int_\Omega \Psi\left(\tilde{\bfepsilon}(\bfx)\right) \dd V
	 + \eta^2 \int_\Omega \Psi\left(\hat{\bfepsilon}(\bfx)\right) \dd V\bigg] \\
	 & = \int_\Omega \Psi\left(\tilde{\bfepsilon}(\bfx)\right) \dd V
	 	- \frac{\eta^2}{(1+\eta)^2}\bigg[ \int_\Omega \Psi\left(\tilde{\bfepsilon}(\bfx)\right) \dd V
	 		- \int_\Omega \Psi\left(\hat{\bfepsilon}(\bfx)\right) \dd V
	 		\bigg].
\end{split}
\ee
Due to assumption~\eqref{eq:assumption}, we know that the final term in brackets is positive and therefore
\be
	W_\text{initial} < \int_\Omega \Psi\left(\tilde{\bfepsilon}(\bfx)\right) \dd V.
\ee
Since the energy inside $\Omega$ cannot exceed its initial value $W_\text{initial}$ for reasons of energy conservation, the energy can never approach the value $\int_\Omega \Psi(\tilde{\bfepsilon}(\bfx))\,\dd V$ so that, if \eqref{eq:assumption} holds,  $\tilde{\bfu}(\bfx)$ cannot be the solution as $t\to\infty$. Hence, if $\tilde{\bfu}(\bfx)$ is a stable solution of linear momentum balance, it must be the minimizer of~\eqref{eq:infcondition}. We note that we did not constrain $\dsC(\bfx)$ to be positive-definite at any point in our proof. Therefore, if in the presence of negative stiffness in a heterogeneous solid a stable equilibrium solution to the Dirichlet problem exists, it must be the minimizer of~\eqref{eq:infcondition}.


\subsection{Sufficiency of the stability conditions for the Dirichlet problem}
\label{sec:Sufficiency}

We showed above that a necessary condition for stability is that the energy is minimized. 
Let us demonstrate in an elastodynamic setting, ignoring viscoelasticity, that this is indeed a sufficient condition of stability if the energy 
still has a minimum when we perturb the elasticity tensor $\dsC(\bfx)$ by subtracting a small constant tensor
\be 
	\delta\dsC_{ijkl}=\frac{\eta}{2}(\delta_{ik}\delta_{jl}+\delta_{il}\delta_{jk})
\ee
from it, where $\eta$ is a small parameter. If $\dsC(\bfx)$ is isotropic, this perturbation corresponds to subtracting a small constant $\eta$ from the shear modulus $\mu(\bfx)$ while leaving the Lam\'{e} modulus $\lambda(\bfx)$ unchanged. Suppose initially at time $t=0$ we begin with a small perturbation $\delta\bfu(\bfx)$ of the solution $\tilde{\bfu}(\bfx)$ which minimizes the energy and no energy enters the domain. Initially at time $t=0$ we could also have a small velocity $\delta\dot\bfu(\bfx)$. Let $\tilde{\bfu}(\bfx)+\delta\bfu(\bfx,t)$ be the displacement field at times $t>0$. We assume that $\delta\bfu(\bfx,t)$ at the boundary $\bfx\in\partial\Omega$ remains zero for all times and that the density $\rho(\bfx)$ is bounded below by some constant $\rho_0>0$. The total energy, i.e.\ the sum of elastic and kinetic parts, equals the minimum energy $\tilde{W}=W[\tilde{\bfu}]$ plus a small perturbation $\delta W$ (due to $\delta\bfu(\bfx)$ and $\delta\dot\bfu(\bfx)$) and must be conserved, i.e. $W=\tilde{W}+\delta W$ is constant. We want to show that $\delta\bfu(\bfx,t)$ and the velocity $\delta\dot\bfu(\bfx,t)=\partial\delta\bfu(\bfx,t)/\partial t$ remain small in an $L_2$ sense for all times. Since the elastic part cannot be less than the minimum $\tilde{W}$, we conclude that the kinetic part must be at most the small perturbation $\delta W$, implying 
\be\label{eq:FirstInequality}
	\frac{\rho_0}{2}\int_\Omega(\delta\dot\bfu(\bfx,t))^2~\dd V\leq\delta W, 
\ee
and the elastic energy is at most $\tilde{W}$+$\delta W$, giving
\be
\begin{split}
	\delta W \geq W[\tilde{\bfu}+\delta\bfu(\bfx,t)]-W[\tilde{\bfu}]
	& = \frac{1}{2}\int_{\partial\Omega_u}\delta\bfepsilon(\bfx,t)\cdot\dsC(\bfx)\,\delta\bfepsilon(\bfx,t)~\dd V \\
	& = \eta \int_\Omega \delta\bfepsilon\cdot\delta\bfepsilon~\dd V 
		+ \frac{1}{2}\int_{\partial\Omega_u}\delta\bfepsilon\cdot[\dsC-\delta\dsC]\,\delta\bfepsilon~\dd V \\
	& \geq \eta \int_\Omega \delta\bfepsilon\cdot\delta\bfepsilon~\dd V.
\end{split}
\ee
Here we have used the fact that $\tilde{\bfu}$ satisfies the equilibrium equation, and that
\be 
	\int_{\partial\Omega_u}\delta\bfepsilon(\bfx,t)\cdot[\dsC(\bfx)-\delta\dsC]\,\delta\bfepsilon(\bfx,t)~\dd V\geq 0,
\ee
which is a necessary condition for a mimimum to exist when the elasticity tensor is $\dsC(\bfx)-\delta\dsC$. 

Following~\citet{EricksenToupin1956}, we can use the fact that $\delta\bfu(\bfx)=\bfnull$ on $\partial\Omega$ (with outward unit normal $\bfn$) by writing
\be
	0 = \int_{\partial\Omega} \left(\delta u_i\,\delta u_{j,j} - \delta u_j\,\delta u_{i,j}\right)n_i \,\dd S
	= \int_\Omega \left(\delta u_i\,\delta u_{j,j} - \delta u_j\,\delta u_{i,j}\right)_{,i} \,\dd V
	= \int_\Omega \left((\delta u_{i,i})^2 - \delta u_{j,i}\,\delta u_{i,j}\right) \,\dd V
\ee
so that
\be\label{eq:Toupin}
	\int_\Omega \delta u_{i,j}\,\delta u_{j,i} \,\dd V
	= \int_\Omega (\delta u_{i,i})^2\,\dd V \geq 0
\ee
and therefore
\be 
	\int_\Omega \delta\bfepsilon\cdot\delta\bfepsilon~\dd V
	= \frac{1}{2}\int_\Omega \left(\delta u_{i,j}\delta u_{i,j}+\delta u_{i,j}\delta u_{j,i}\right)\dd V
	\geq \frac{1}{2}\int_\Omega \delta u_{i,j} \delta u_{i,j}~\dd V.
\ee
Finally, using Poincar\'e's inequality there exists a constant $C_\Omega>0$ such that
\be 
	\int_\Omega \delta\bfu_{i,j}\cdot\delta\bfu_{i,j}~\dd V\geq C_\Omega\int_\Omega \delta\bfu_{i}\cdot\delta\bfu_{i}~\dd V,
\ee
which allows us to conclude that
\be \label{eq:SecondInequality}
	\int_\Omega \delta\bfu_{i}\cdot\delta\bfu_{i}~\dd V\leq \frac{2\,\delta W}{\eta\,C_\Omega}.
\ee
From the inequalities~\eqref{eq:FirstInequality} and~\eqref{eq:SecondInequality} it is evident that $\delta\bfu(\bfx,t)$ and $\delta\dot\bfu(\bfx,t)$ remain small in an $L_2$ sense for all times. Presumably, if we were to add viscoelasticity the displacement $\delta\bfu(\bfx,t)$ would damp to zero as $t\to\infty$. 


\subsection{Neumann problem: natural boundary conditions}

\begin{figure}[t]
	\centering
		\includegraphics[width=0.9\textwidth]{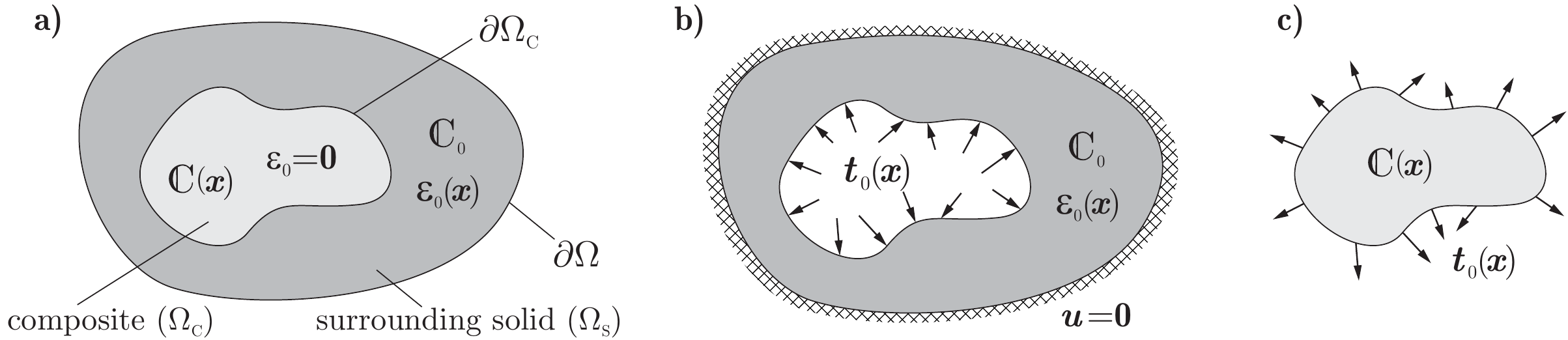}
	\caption{Neumann problem: a) inhomogeneous body embedded in a surrounding solid, b) tractions $\bft_0(\bfx)$ and vanishing outer displacements define strains $\bfepsilon_0(\bfx)$ in the surrounding solid, c) equivalent Neumann boundary value problem for the inhomogeneous body with surface tractions $\bft_0(\bfx)$.}
	\label{fig:Neumann}
\end{figure}

As shown in Fig.~\ref{fig:Neumann}a, we consider an inhomogeneous body $\Omega_\text{C}$ containing a composite material made of linear elastic constituents, which is completely embedded in and perfectly bonded to another linear elastic surrounding solid $\Omega_\text{S}$ (with positive-definite and spatially constant elastic moduli $\dsC_0$), i.e. 
\be
	\dsC(\bfx) = \begin{cases}
		\dsC(\bfx), & \text{if} \ \bfx\in\Omega_\text{C},\\
		\dsC_0, & \text{if} \ \bfx\in\Omega_\text{S}.
	\end{cases}
\ee
We assume that displacements vanish on the outer surface, i.e.\ $\bfu=\bfnull$ on $\partial\Omega$, where $\Omega=\Omega_\text{C}\cup\Omega_\text{S}$ denotes the entire solid. We assume that eigenstrains $\bfepsilon_0(\bfx)$ act within the surrounding solid but not within the inhomogeneous body:
\be
	\bfepsilon_0(\bfx) = \begin{cases}
		\bfnull, & \text{if} \ \bfx\in\Omega_\text{C},\\
		\bfepsilon_0(\bfx), & \text{if} \ \bfx\in\Omega_\text{S}.
	\end{cases}
\ee
Let us construct the eigenstrains $\bfepsilon_0(\bfx)$ in the following way: consider the surrounding solid with the inhomogeneous body removed. Next, apply tractions $\bft_0(\bfx)$ on the inner boundary and apply zero displacements $\bfu_0=\bfnull$ on the outer boundary, as shown in Fig.~\ref{fig:Neumann}b. The resulting strain field that balances the applied tractions in equilibrium will be taken as our eigenstrain $\bfepsilon_0(\bfx)$, so that because of linear momentum balance we have
\be
	\divv\bfsigma_0(\bfx) = \divv \left[\dsC_0\,\bfepsilon_0(\bfx)\right] = 0 \quad \text{in} \ \Omega_\text{S}
\ee
and $\bft_0(\bfx)=\bfsigma_0(\bfx)\bfn_{\partial\Omega_\text{S}}(\bfx)$ on the inner boundary $\partial\Omega_\text{C}$ with unit normal $\bfn_{\partial\Omega_\text{S}}(\bfx)$ pointing outward from $\partial\Omega_\text{S}$.

Going back to the system consisting of inhomogeneous body and surrounding elastic medium, assume $\tilde{\bfu}(\bfx)$ is the displacement field corresponding to a solution of the elasticity equation (linear momentum balance) in the presence of eigenstrains $\bfepsilon_0(\bfx)$ as determined above with essential boundary conditions $\tilde{\bfu}(\bfx)=\bfnull$ on the outer boundary $\partial\Omega$. From Section~\ref{sec:DirichletProblem} we know that a necessary condition for stability is that $\tilde{\bfu}(\bfx)$ minimizes the total elastic energy
\be\label{eq:DirichletCondEigen}
	W = \inf_{\Large\bfu(\bfx) \atop \bfu(\bfx)=\bfnull \ \text{on} \ \partial\Omega}
		\int_\Omega \frac{1}{2} \left[\bfepsilon(\bfx)-\bfepsilon_0(\bfx)\right]\cdot \dsC(\bfx) \,
		\left[\bfepsilon(\bfx)-\bfepsilon_0(\bfx)\right] \,\dd V.
\ee
For the given problem, we can expand the energy as follows:
\be\label{eq:EnergySplitting}
\begin{split}
	\int_\Omega \frac{1}{2} & \left[\bfepsilon(\bfx)-\bfepsilon_0(\bfx)\right]\cdot \dsC(\bfx) \,
		\left[\bfepsilon(\bfx)-\bfepsilon_0(\bfx)\right] \,\dd V\\ 
	& \qquad = \int_\Omega \frac{1}{2} \bfepsilon(\bfx)\cdot \dsC(\bfx) \, \bfepsilon(\bfx)\,\dd V
		- \int_{\Omega_\text{S}} \bfepsilon(\bfx)\cdot \dsC_0 \, \bfepsilon_0(\bfx)\,\dd V
		+ \int_{\Omega_\text{S}} \frac{1}{2} \bfepsilon_0\cdot \dsC_0 \, \bfepsilon_0(\bfx)\,\dd V,
\end{split}
\ee
where the last term is constant and independent of the displacement field. The second term can further be reduced by using $\bfsigma_0(\bfx)=\dsC_0\,\bfepsilon_0(\bfx)$, which yields
\be
\begin{split}
	\int_{\Omega_\text{S}} \bfepsilon(\bfx)\cdot \dsC_0 \, \bfepsilon_0(\bfx)\,\dd V 
	& = \int_{\Omega_\text{S}} \grad\bfu(\bfx)\cdot \bfsigma_0(\bfx) \, \dd V
		= \int_{\partial\Omega_\text{S}} \bfsigma_0(\bfx)\bfn_{\partial\Omega_\text{S}}(\bfx)\cdot\bfu(\bfx)\,\dd S
			- \int_{\Omega_\text{S}} \divv\bfsigma_0(\bfx)\cdot\bfu(\bfx)\,\dd S\\
	&	= \int_{\partial\Omega_\text{C}} \bft_0(\bfx)\cdot\bfu(\bfx)\,\dd S
			+ \int_{\partial\Omega} \bfsigma_0(\bfx)\bfn_{\partial\Omega_\text{S}}(\bfx)\cdot\bfu(\bfx)\,\dd S
			- \int_{\Omega_\text{S}} \divv\bfsigma_0(\bfx)\cdot\bfu(\bfx)\,\dd S,
\end{split}
\ee
Using that $\bfu(\bfx)=\bfnull$ on $\partial\Omega$ and $\divv\bfsigma_0(\bfx)=\bfnull$ inside $\Omega_\text{S}$ as well as $\bft_0(\bfx)=\bfsigma_0(\bfx)\bfn_{\partial\Omega_\text{S}}$, we thus obtain
\be
	\int_{\Omega_\text{S}} \bfepsilon(\bfx)\cdot \dsC(\bfx) \, \bfepsilon_0\,\dd V 
	= \int_{\partial\Omega_\text{C}} \bft_0(\bfx)\cdot\bfu(\bfx)\,\dd S.
\ee
Moreover, the first term in~\eqref{eq:EnergySplitting} can be decomposed into strain energy stored in the inhomogeneous body and in the surrounding solid, i.e.
\be\label{eq:SplitVolumeIntegral}
	\int_\Omega \frac{1}{2} \bfepsilon(\bfx)\cdot \dsC(\bfx) \, \bfepsilon(\bfx)\,\dd V
	= \int_{\Omega_{C}} \frac{1}{2} \bfepsilon(\bfx)\cdot \dsC(\bfx) \, \bfepsilon(\bfx)\,\dd V
		+ \int_{\Omega_{S}} \frac{1}{2} \bfepsilon(\bfx)\cdot \dsC_0 \, \bfepsilon(\bfx)\,\dd V.
\ee
Altogether we thus have
\be
\begin{split}
	& \int_\Omega \frac{1}{2} \left[\bfepsilon(\bfx)-\bfepsilon_0(\bfx)\right]\cdot \dsC(\bfx) \,
		\left[\bfepsilon(\bfx)-\bfepsilon_0(\bfx)\right] \,\dd V\\
	& \qquad = \int_{\Omega_{C}} \frac{1}{2} \bfepsilon(\bfx)\cdot \dsC(\bfx) \, \bfepsilon(\bfx)\,\dd V
		+ \int_{\Omega_{S}} \frac{1}{2} \bfepsilon(\bfx)\cdot \dsC_0 \, \bfepsilon(\bfx)\,\dd V
		- \int_{\partial\Omega_{C}} \bft_0(\bfx)\cdot \bfu(\bfx)\,\dd S + W_0,
\end{split}
\ee
where 
\be
	W_0 = \int_{\Omega_\text{S}} \frac{1}{2} \,\bfepsilon_0(\bfx)\cdot \dsC_0 \, \bfepsilon_0(\bfx)\,\dd V = \text{const.} >0.
\ee
Recall that we determined the eigenstresses $\bfsigma_0(\bfx)=\dsC_0\,\bfepsilon_0(\bfx)$ from the application of tractions $\bft_0(\bfx)$ on the interface and vanishing displacements on $\partial\Omega$. Now, keep tractions $\bft_0(\bfx)$ constant and consider a scaling of the elastic moduli in the surrounding solid of the following form:
\be
	\dsC_0=\alpha\,\overline{\dsC}_0 \quad \Rightarrow \quad 
	\bfepsilon_0 = \dsC^{-1}_0\bfsigma_0 = \frac{\overline{\dsC}_0^{-1}\bfsigma_0 }{\alpha}
	= \frac{\overline{\bfepsilon_0}}{\alpha}
\ee
with some $\alpha>0$. This gives
\be
\begin{split}
	& \int_\Omega \frac{1}{2} \left[\bfepsilon(\bfx)-\bfepsilon_0(\bfx)\right]\cdot \dsC(\bfx) \,
		\left[\bfepsilon(\bfx)-\bfepsilon_0(\bfx)\right] \,\dd V\\
	& \qquad = \int_{\Omega_{C}} \frac{1}{2} \bfepsilon(\bfx)\cdot \dsC(\bfx) \, \bfepsilon(\bfx)\,\dd V
		+ \alpha\,\int_{\Omega_{S}} \frac{1}{2} \bfepsilon(\bfx)\cdot \overline{\dsC}_0 \, \bfepsilon(\bfx)\,\dd V
		- \int_{\partial\Omega_{C}} \bft_0(\bfx)\cdot \bfu(\bfx)\,\dd S + \frac{\overline{W}_0}{\alpha}.
\end{split}
\ee
Here, we may choose $\alpha$ arbitrarily small so that the second term vanishes (and the final term has no effect even though it grows in an unbounded manner since it is independent of the displacement field). To a good approximation when $\alpha$ is extremely small $\tilde{\bfu}(\bfx)$ in $\Omega_\text{C}$ is the approximate minimizer of
\be\label{eq:NeumannVariationalP}
	W = \inf_{\bfu(\bfx)} \left\{\int_{\Omega_\text{C}} \frac{1}{2} \bfepsilon(\bfx)\cdot \dsC(\bfx) \, \bfepsilon(\bfx)\,\dd V
		- \int_{\partial\Omega_{C}} \bft_0(\bfx)\cdot \bfu(\bfx)\,\dd S \right\}
\ee
and $\tilde{\bfu}(\bfx)$ in $\Omega_\text{S}$ is approximately the minimizer of
\be
	\inf_{\bfu(\bfx)} \int_{\Omega_\text{S}} \frac{1}{2} \bfepsilon(\bfx)\cdot \overline{\dsC}_0 \, \bfepsilon(\bfx)\,\dd V
\ee
subject to the constraint that $\bfu(\bfx)=\bfnull$ on the outer boundary $\partial\Omega$ and that, on the boundary $\partial\Omega_\text{C}$, $\bfu(\bfx)$ equals the displacement that minimizes~\eqref{eq:NeumannVariationalP} -- to ensure that this approximate solution for $\tilde{\bfu}(\bfx)$ is continuous across $\partial\Omega_\text{C}$. Furthermore, when $\alpha$ is infinitesimal the stress in $\Omega_\text{S}$,
\be
	\tilde{\bfsigma}(\bfx) = \dsC_0\left[\tilde{\bfepsilon}(\bfx) - \bfepsilon_0(\bfx)\right]
	= \alpha\,\overline{\dsC}_0\,\tilde{\bfepsilon}(\bfx) - \bfsigma_0(\bfx)
\ee
approaches $-\bfsigma_0(\bfx)$ and the tractions on the interface (seen from $\Omega_\text{S}$) follow as $\bft_{\partial\Omega_\text{S}}(\bfx)=-\bfsigma_0(\bfx)\bfn_{\partial\Omega_\text{S}}(\bfx)=-\bft_0(\bfx)$. Therefore, due to balance of tractions (i.e.\ $\bft_{\partial\Omega_\text{C}}=-\bft_{\partial\Omega_\text{S}}$), $\bft_0(\bfx)$ can be identified with the traction acting on the surface $\partial\Omega_\text{C}$, see Fig.~\ref{fig:Neumann}c. In summary, from~\eqref{eq:DirichletCondEigen} it follows that a necessary condition of stability with traction boundary conditions is that the displacement field $\tilde{\bfu}(\bfx)$ is a minimizer of the total potential energy~\eqref{eq:NeumannVariationalP} with $\bft_0$ begin the tractions applied to the surface $\partial\Omega_\text{C}$. Notice that the opposite limit of letting $\alpha$ become infinitely large recovers the Dirichlet boundary value problem.


The dual variational principle 
\be\label{eq:DualPrinc}
	\overline{W} = \inf_{\Large\bfsigma(\bfx) \atop \bfsigma(\bfx)\bfn(\bfx)=\bft(\bfx) \ \text{on} \ \partial\Omega}
		\int_\Omega \frac{1}{2}\,\bfsigma(\bfx)\cdot \dsC^{-1}(\bfx) \,\bfsigma(\bfx) \,\dd V,
\ee
which has been used for traction boundary problems and, among others, yields the Reuss lower bound on the effective moduli of inhomogeneous solids and composites does \emph{not} apply. This will be shown by the aid of an instructive example in Section~\ref{sec:dual}.


\subsection{Stability conditions for homogeneous isotropic linear elastic solids}
\label{sec:HomogeneousSolids}

Stability conditions for homogeneous solids (for with $\dsC(\bfx)=\dsC=$ const.) can be obtained from the variational principles shown above. For the Dirichlet problem, principle~\eqref{eq:infcondition} can be rephrased by taking variations as
\be
	\delta^2 W = \frac{1}{2}\int_\Omega \dsC_{ijkl}(\bfx)\,\delta u_{i,j}(\bfx)\,\delta u_{k,l}(\bfx)\,\dd V \geq 0 \qquad \forall \quad \delta\bfu(\bfx) \quad \text{with} \quad \delta\bfu(\bfx)=\bfnull \quad \text{on} \ \partial\Omega,
\ee
For a homogeneous solid with spatially constant elastic moduli, we conclude that
\be
	\dsC_{ijkl}\, \int_\Omega \delta u_{i,j}(\bfx)\,\delta u_{k,l}(\bfx)\,\dd V \geq 0  \qquad \forall \quad \delta\bfu(\bfx) \quad \text{with} \quad \delta\bfu(\bfx)=\bfnull \quad \text{on} \ \partial\Omega.
\ee
In the special case of an isotropic solid with Lam\'{e} moduli $\lambda$ and $\mu$ ($\mu$ being the shear modulus), we have
\be
	\dsC_{ijkl} = \lambda\,\delta_{ij}\,\delta_{kl} + \mu \left(\delta_{ik}\,\delta_{jl} + \delta_{il}\,\delta_{jk}\right)
\ee
so that the stability condition becomes
\be
	\lambda\, \langle \delta u_{i,i}^2(\bfx)\rangle + \mu \left\langle\,\delta u_{i,j}(\bfx)\,\delta u_{i,j}(\bfx) 
	+ \delta u_{i,j}(\bfx)\,\delta u_{j,i}(\bfx)\,\right\rangle \geq 0  \qquad \forall \quad \delta\bfu(\bfx) \quad \text{with} \quad \delta\bfu(\bfx)=\bfnull \quad \text{on} \ \partial\Omega,
\ee
where $\langle\cdot\rangle = \frac{1}{V}\int_\Omega(\cdot)\,\dd V$. Introducing the infinitesimal rotation tensor $\omega_{ij} = \frac{1}{2}(u_{i,j} - u_{j,i})$ gives
\be
	\omega_{ij}\,\omega_{ij} = \frac{1}{4}\left(u_{i,j} - u_{j,i}\right)\left(u_{i,j} - u_{j,i}\right)
	= \frac{1}{2}\left(u_{i,j}\,u_{i,j} - u_{i,j}\,u_{j,i}\right),
\ee
which, using~\eqref{eq:Toupin}, ultimately leads to 
\be
\begin{split}
	\lambda\, \langle \delta u_{i,i}^2\rangle + \mu \,\langle\,\delta u_{i,j}\,\delta u_{i,j} + \delta u_{i,j}\,\delta u_{j,i}\rangle
	& = (\lambda + \mu) \,\langle \delta u_{i,i}^2\rangle + \mu \,\langle 2\,\delta\omega_{ij}\,\delta\omega_{ij} 
		+ \delta u_{i,j}\,\delta u_{j,i}\rangle\\
	& = (\lambda + 2\mu) \,\langle \delta u_{i,i}^2\rangle + 2\mu\, \langle \delta\omega_{ij}\,\delta\omega_{ij}\rangle \geq 0.
\end{split}
\ee
Because we can make a large twist inside the body and make $\langle\delta\omega_{ij}\delta\omega_{ij}\rangle$ arbitrarily large while keeping $\langle\delta u_{i,i}^2\rangle$ bounded, or alternatively make a large local compression and make $\langle\delta u_{i,i}^2\rangle$ arbitrarily large while keeping $\langle\delta\omega_{ij}\delta\omega_{ij}\rangle$ bounded, a necessary and sufficient condition of stability is given by the well-known conditions of strong ellipticity \citep{EricksenToupin1956,Hill1957}
\be\label{eq:StrongEllipticityIsotropic}
	\mu>0 \qquad \text{and} \qquad \lambda+2\,\mu>0.
\ee
Note that conditions~\eqref{eq:StrongEllipticityIsotropic} agree with \citeauthor{Hadamard1903}'s \citeyearpar{Hadamard1903} necessary conditions of pointwise stability in elastic media, which ensure real-valued wave speeds~\citep{Kelvin1888}.

For the Neumann problem, the variational principle~\eqref{eq:NeumannVariationalP} holds, from which we can obtain stability conditions for homogeneous solids again by considering the second variation:
\be
	\delta^2 W = \int_\Omega \dsC_{ijkl}\,\delta u_{i,j}(\bfx)\,\delta u_{k,l}(\bfx)\,\dd V \geq 0 \qquad \forall \quad \delta\bfu(\bfx)
	\neq\bfnull.
\ee
We may decompose the displacement gradient into its volumetric and deviatoric contributions, i.e.\ $u_{i,j} = u_{i,j}^\text{vol} + u_{i,j}^\text{dev}$ with $u_{i,j}^\text{dev} = u_{i,j} - \frac{1}{3}u_{k,k}\delta_{ij}$ and note that $u_{i,i}^\text{dev}=0$. For homogeneous, isotropic, linear elastic solids we have
\be
\begin{split}
	\dsC_{ijkl}\,\left\langle \delta u_{i,j}(\bfx)\,\delta u_{k,l}(\bfx)\right\rangle & =
		\lambda\, \left\langle \delta u_{i,i}^2(\bfx)\rangle + \mu \,\langle\,\delta u_{i,j}(\bfx)\,\delta u_{i,j}(\bfx) 
			+ \delta u_{i,j}(\bfx)\,\delta u_{j,i}(\bfx)\right\rangle\\
	& = \left(\lambda+\frac{2}{3}\mu\right) \left\langle \delta u_{k,k}^2(\bfx)\right\rangle + 2\,\mu\,\left\langle 	
		\delta\varepsilon_{ij}(\bfx)^\text{dev}\,\delta\varepsilon_{ij}(\bfx)^\text{dev}\right\rangle,
\end{split}
\ee
and therefore the stability condition becomes
\be
	\left(\lambda+\frac{2}{3}\mu\right) \langle \delta \varepsilon_{kk}^2\rangle + 2\mu\,\langle 	
		\delta\varepsilon_{ij}^\text{dev}\,\delta\varepsilon_{ij}^\text{dev}\rangle \geq 0 
	\qquad \forall \quad \delta\bfu(\bfx)\neq\bfnull.
\ee
Because we can take the strain $\varepsilon_{ij}(\bfx)$ to be constant throughout the body, with either vanishing volumetric strain, or vanishing deviatoric strain, this implies the necessary and sufficient conditions of stability for the Neumann problem,
\be\label{eq:PosDefIsotropic}
	\mu > 0 \qquad \text{and} \qquad \kappa = \lambda+\frac{2}{3}\mu > 0,
\ee
which are the well-known conditions of positive-definiteness of the elastic modulus tensor~\citep{Kirchhoff1859} with $\kappa$ denoting the bulk modulus.

In case of homogeneous anisotropic linear elastic solids, the same variational principles apply and the existence of a unique minimizer requires quasiconvexity of the total potential energy, see e.g.~\citep{KnopsStuart1984}. The resultant necessary and sufficient condition of stability is positive-definiteness of the elastic modulus tensor, i.e.
\be
	\bfepsilon \cdot\dsC\,\bfepsilon > 0 \qquad \text{for all symmetric second-order tensors} \ \bfepsilon\neq 0,
\ee
which for isotropy automatically reduces to~\eqref{eq:PosDefIsotropic}. In case of pure displacement boundary conditions, the necessary and sufficient condition of stability is strong ellipticity of the elastic modulus tensor~\citep{Hadamard1903}, i.e.
\be
	(\bfa\otimes\bfn)\cdot\dsC\,(\bfa\otimes\bfn) > 0 \qquad \text{for all vectors} \ \bfa,\bfn\neq 0.
\ee
For isotropy this reduces to~\eqref{eq:StrongEllipticityIsotropic}.


\begin{figure}[t]
	\centering
		\includegraphics[width=0.70\textwidth]{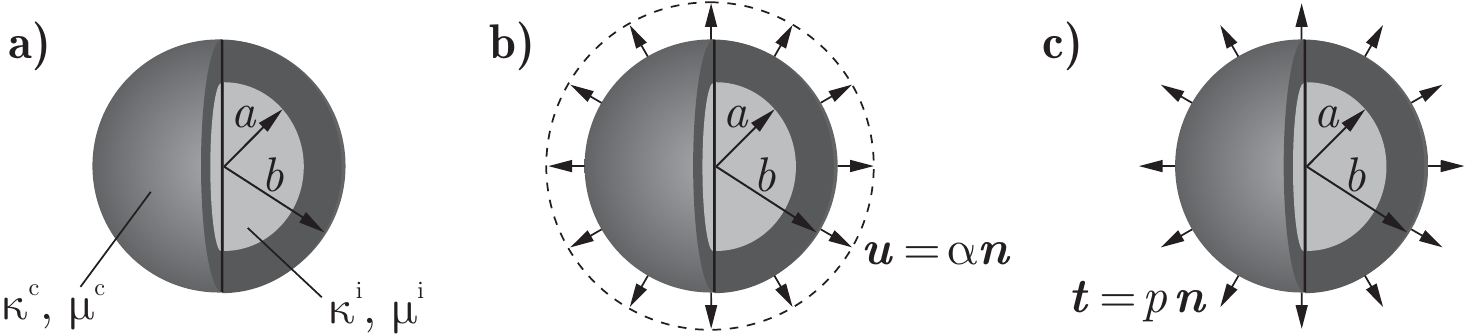}
	\caption{a) Example of a coated spherical inclusion with b) Dirichlet and c) Neumann boundary conditions.}
	\label{fig:spheres}
\end{figure}

\section{Example: coated spherical inclusion}
\label{sec:CoatedSphere}

Consider a two-phase body consisting of a homogeneous, isotropic, linear elastic spherical particle (radius $a$, elastic moduli $\mu^\text{i}$ and $\kappa^\text{i}$) coated by and perfectly bonded to a concentric coating of outer radius $b$ and of a different homogeneous, isotropic, linear elastic material (elastic moduli $\mu^\text{c}$ and $\kappa^\text{c}$) as schematically shown in Fig.~\ref{fig:spheres}a. The system was studied before to derive effective properties and stability conditions, see e.g.~\citep{LakesDrugan2002,KochmannDrugan2012, WojnarKochmann2013a}. The same example of a two-phase solid will be used here to demonstrate the applicability and inapplicability of the standard variational principles in the presence of negative-stiffness phases and their relations to the conditions of overall stability. For simplicity, we assume radial symmetry and choose the boundary conditions accordingly.

\subsection{Dirichlet boundary value problem}

We impose a radial displacement field $\bfu(b) = \alpha\,\bfn$ across the entire outer surface with outward unit normal $\bfn$ and constant $\alpha\in\Rset$, which results in radial displacements in the inclusion (superscript \emph{i}) and in the coating (superscript \emph{c}) of Lam\'{e}'s type, viz.\
\begin{subequations}
\begin{align}
	\bfu^\text{i}(r)& = u_r^\text{i}(r)\,\bfe_r, \qquad u_r^\text{i}(r) = A\,r\\
	\bfu^\text{c}(r)& = u_r^\text{c}(r)\,\bfe_r, \qquad u_r^\text{c}(r) = B\,r + \frac{C}{r^2}
\end{align}\label{eq:Displacements}%
\end{subequations}
in spherical coordinates $(r,\theta,\varphi)$. In static equilibrium, constants $A$, $B$ and $C$ are determined by application of the boundary and continuity conditions 
\be
	u_r^\text{c}(b)=\alpha , \qquad
	\sigma_{rr}^\text{i}(a)=\sigma_{rr}^\text{c}(a), \qquad
	u_r^\text{i}(a)=u_r^\text{c}(a).
\ee
The stress components $\sigma_{rr}$ are determined from the displacements~\eqref{eq:Displacements} by application of the strain-displacement relation $\bfepsilon = \frac{1}{2}(\grad\bfu+ \grad\bfu\T)$ and Hooke's law for isotropic elasticity, $\bfsigma = \dsC\,\bfepsilon$. The equilibrium solution is then given by~\eqref{eq:Displacements} with
\be\label{eq:Displacements1}
	A = \frac{3 \kappa^\text{c} + 4\mu^\text{c}}{3 \kappa^\text{i} + 4 \mu^\text{c} + 3(a/b)^3(\kappa^\text{c}-\kappa^\text{i})}
	\frac{\alpha}{b},\quad
	B = \frac{3 \kappa^\text{i} + 4\mu^\text{c}}{3 \kappa^\text{i} + 4 \mu^\text{c} + 3(a/b)^3(\kappa^\text{c}-\kappa^\text{i})}
	\frac{\alpha}{b},\quad
	C = \frac{3 (\kappa^\text{c}-\kappa^\text{i})a^3}{3 \kappa^\text{i} + 4 \mu^\text{c}+3(a/b)^3(\kappa^\text{c}-\kappa^\text{i})}
	\frac{\alpha}{b}
\ee
and therefore the pressure on the outer surface follows as
\be
	\sigma_{rr}^\text{c}(b) = \frac{3 \alpha}{b} \frac{\kappa^\text{c}\left(3\kappa^\text{i}+4\mu^\text{c} \right)
		+	4\,(a/b)^3\left(\kappa^\text{i}-\kappa^\text{c}\right)\mu^\text{c}}
		{3 \kappa^\text{i} + 4 \mu^\text{c} + 3(a/b)^3(\kappa^\text{c}-\kappa^\text{i})}.
\ee
Note that we can also obtain the effective bulk modulus of the associated two-phase assemblage of coated spheres for the Dirichlet problem via~\citep{Hashin1962}
\be\label{eq:EffBulkSphere}
	\kappa_*^\text{D} = \frac{1}{3}\frac{\langle\tr\bfsigma\rangle}{\langle\tr\bfepsilon\rangle} = \frac{\sigma_{rr}^\text{c}(b)\,b}{3\,u_r^\text{c}(b)}
	= \frac{b}{3\alpha}\,\sigma_{rr}^\text{c}(b)
	= \frac{\kappa^\text{c}\left(3\kappa^\text{i}+4\mu^\text{c} \right)
		+	4\,(a/b)^3\left(\kappa^\text{i}-\kappa^\text{c}\right)\mu^\text{c}}
		{3 \kappa^\text{i} + 4 \mu^\text{c} + 3(a/b)^3(\kappa^\text{c}-\kappa^\text{i})}
\ee
with volume averages $\langle\cdot\rangle = \frac{1}{V}\int_\Omega (\cdot)\,\dd V$. Therefore, an infinite effective bulk modulus is predicted when~\citep{LakesDrugan2002}
\be\label{eq:kappaMin}
	\kappa^\text{i}\rightharpoondown - \frac{3\, a^3 \kappa^\text{c}+4\, b^3 \mu^\text{c}}{3 \left(b^3-a^3\right)} = \kappa^\text{i}_{\infty},
\ee
i.e.\ when $\kappa^\text{i}$ approaches the root of the denominator in~\eqref{eq:EffBulkSphere} from below. Also, with decreasing inclusion bulk modulus, the effective bulk modulus first goes to zero when the numerator in~\eqref{eq:EffBulkSphere} vanishes, i.e.\ when
\be
	\kappa^\text{i} = - \frac{4 \left(b^3-a^3\right) \kappa^\text{c} \mu^\text{c}}{3 b^3 \kappa^\text{c}+4 a^3 \mu^\text{c}}
	= \kappa^\text{i}_0.
\ee
Simple algebraic manipulations show that for all combinations of strongly-elliptic elastic moduli (required for pointwise stability) and radii $b>a$ we have that $\kappa^\text{i}_{\infty}<\kappa^\text{i}_0$.

Next, let us verify the applicability of variational principle~\eqref{eq:infcondition} derived above for the Dirichlet problem. To this end, we introduce a space of displacement field solutions which are continuous inside the solid and satisfy the boundary condition, such that the space of solutions contains the equilibrium solution~\eqref{eq:Displacements} with~\eqref{eq:Displacements1}. For example, consider a displacement field $\tilde{\bfu}(r)$ identical to~\eqref{eq:Displacements} whose coefficients $\tilde{A}$, $\tilde{B}$ and $\tilde{C}$ are determined by enforcing 
\be
	\tilde{u}_r^\text{c}(b)=\alpha, \qquad
	\tilde{u}_r^\text{i}(a)=u_a, \qquad
	\tilde{u}_r^\text{c}(a)=u_a
\ee
for some interface displacement $u_a$, which results in
\be
	\tilde{A} = \frac{u_a}{a},\quad 
	\tilde{B} = \frac{\alpha\,b^2 - u_a\,a^2}{b^3-a^3},\quad
	\tilde{C} = \frac{(u_a\,b-\alpha\,a)a^2b^2}{b^3-a^3}
\ee
with an unknown $u_a$. Note that the correct equilibrium solution~\eqref{eq:Displacements1} is contained herein and attained when choosing
\be\label{eq:ua}
	u_a = A\,a = \frac{3 \kappa^\text{c} + 4\mu^\text{c}}{3 \kappa^\text{i} + 4 \mu^\text{c} + 3(a/b)^3(\kappa^\text{c}-\kappa^\text{i})}
	\frac{\alpha\,a}{b}.
\ee
The total energy of the two-phase body $\Omega$ in the absence of eigenstrains is given by
\be
\begin{split}
	\tilde{W} & = \frac{1}{2}\int_\Omega\tilde{\bfepsilon}\cdot\dsC\,\tilde{\bfepsilon}\,\dd V 
		= \frac{1}{2}\int_\Omega\left[\kappa\,(\tr\tilde{\bfepsilon})^2 
			+ \mu\,\tilde{\bfepsilon}_\text{dev}\cdot\tilde{\bfepsilon}_\text{dev}\right]\,\dd V\\
		& = \frac{1}{2}\int_0^a \left[\kappa^\text{i}(\tr\tilde{\bfepsilon}^\text{i})^2 
			+ \mu^\text{i}\,\tilde{\bfepsilon}_\text{dev}^\text{i}\cdot\tilde{\bfepsilon}_\text{dev}^\text{i}\right]\,4\pi r^2 \,\dd r
			+ \frac{1}{2}\int_a^b \left[\kappa^\text{c}(\tr\tilde{\bfepsilon}^\text{c})^2 
			+ \mu^\text{c}\,\tilde{\bfepsilon}_\text{dev}^\text{c}\cdot\tilde{\bfepsilon}_\text{dev}^\text{c}\right]\,4\pi r^2 \,\dd r\\
		& = \frac{4 \pi}{b^3-a^3} \,\left[3 a \left(b^3-a^3\right) u_a^2\, \kappa^\text{i}+3 \left(a^2 u_a-b^2 \alpha \right)^2 \kappa^\text{c}
			+ 4\, a\, b \,(b\, u_a -a \,\alpha )^2 \mu^\text{c}\right],
\end{split}
\ee
where $\tilde{\bfepsilon}_\text{dev}=\tilde{\bfepsilon}-\frac{1}{3}(\tr\tilde{\bfepsilon})\id$ is the deviatoric strain tensor. According to~\eqref{eq:infcondition}, the equilibrium solution $\tilde{u}_a$ can be found by minimization:
\be
	\tilde{u}_a = \argmin \tilde{W} \qquad\Rightarrow\qquad
	\frac{\partial \tilde{W}}{\partial u_a} = 0,
\ee
which yields~\eqref{eq:ua}, i.e.\ the correct equilibrium solution. To signal whether this equilibrium solution corresponds to an energy minimum, we note that
\be
	\frac{\partial^2 \tilde{W}}{\partial u_a^2} = 8\, a\, \pi\,\frac{ 3 a^3 (\kappa^\text{i}-\kappa^\text{c}) 
		- b^3 (3 \kappa^\text{i}+4 \mu^\text{c})}{a^3-b^3}
	\begin{cases}
		> 0 & \ \text{if} \quad \kappa^\text{i} > -\cfrac{3\, a^3 \kappa^\text{c}+4\, b^3 \mu^\text{c}}{3 \left(b^3-a^3\right)}
			= \kappa^\text{i}_\infty,\\[0.3cm]
		\leq 0 & \ \text{if} \quad \kappa^\text{i} \leq -\cfrac{3\, a^3 \kappa^\text{c}+4\, b^3 \mu^\text{c}}{3 \left(b^3-a^3\right)}
			= \kappa^\text{i}_\infty.
	\end{cases}
\ee
Consequently, the equilibrium solution may present a stable energy minimum for the assumed variation only if $\kappa^\text{i} > \kappa^\text{i}_{\infty}$ so that a positive-infinite effective bulk modulus cannot be stable, which confirms previous results~\citep{KochmannDrugan2012,WojnarKochmann2013b}. We note that, in addition, pointwise stability requires $\kappa^\text{i} > -\frac{4}{3}\mu^\text{i}$.


\subsection{Neumann boundary value problem}

For the corresponding Neumann boundary value problem, we apply a uniform pressure $p$ to the entire outer surface of the coated sphere and hence apply the boundary and continuity conditions
\be\label{eq:NeumannBCs}
	\sigma_{rr}^\text{c}(b)=p, \qquad
	\sigma_{rr}^\text{i}(a)=\sigma_{rr}^\text{c}(a), \qquad
	u_r^\text{i}(a)=u_r^\text{c}(a).
\ee
The general representation of the displacement field is again given by~\eqref{eq:Displacements}, for which enforcement of~\eqref{eq:NeumannBCs} results in the solution
\be\label{eq:NeumannSolution}
	A = \frac{b^3 p \,(3 \kappa^\text{c}+4 \mu^\text{c})}
		{3D}, \qquad
	B = \frac{b^3 p \,(3 \kappa^\text{i}+4 \mu^\text{c})}
		{3D}, \qquad
	C = \frac{a^3 b^3 p \, (\kappa^\text{c}-\kappa^\text{i})}
		{D}
\ee
with $D=4 a^3 (\kappa^\text{i}-\kappa^\text{c}) \mu^\text{c}+b^3 \kappa^\text{c} (3 \kappa^\text{i}+4 \mu^\text{c})$. The effective bulk modulus of the associated coated-sphere assemblage for the Neumann problem is obtained as
\be
	\kappa_*^\text{N} = \frac{1}{3}\frac{\langle\tr\bfsigma\rangle}{\langle\tr\bfepsilon\rangle} = \frac{p\,b}{3\,u_r^\text{c}(b)}
	= \frac{\kappa^\text{c}\left(3\kappa^\text{i}+4\mu^\text{c} \right)
		+	4\,(a/b)^3\left(\kappa^\text{i}-\kappa^\text{c}\right)\mu^\text{c}}
		{3 \kappa^\text{i} + 4 \mu^\text{c} + 3(a/b)^3(\kappa^\text{c}-\kappa^\text{i})}
	= \kappa_*^\text{D}.
\ee
As can be expected, for the chosen geometry both Dirichlet and Neumann problem yield the same effective structural bulk modulus of the coated-sphere assemblage (and the two displacement field solutions are identical if we choose \mbox{$p=3\,\alpha\,\kappa_*/b$}). This is not the case in general, which is why in subsequent sections we will keep the differentiation.

By analogy with the Dirichlet problem, let us verify the applicability of variational principle~\eqref{eq:DirichletCondEigen} derived above for the Neumann problem. To this end, we introduce a space of displacement field solutions analogously to the previous case. Consider a displacement field $\tilde{\bfu}(r)$ identical to~\eqref{eq:Displacements} whose coefficients $\tilde{A}$, $\tilde{B}$ and $\tilde{C}$ are now determined by 
\be
	\tilde{\sigma}_{rr}^\text{c}(b)=p, \qquad
	\tilde{u}_r^\text{i}(a)=u_a, \qquad
	\tilde{u}_r^\text{c}(a)=u_a
\ee
with an unknown interface displacement $u_a$. These conditions result in
\be
	\tilde{A} = \frac{u_a}{a},\qquad 
	\tilde{B} = \frac{b^3 p+4 a^2 u_a \mu^\text{c}}{3 b^3 \kappa^\text{c}+4 a^3 \mu^\text{c}},\qquad
	\tilde{C} = \frac{a^2 b^3 (3 u_a \kappa^\text{c}-a p)}{3 b^3 \kappa^\text{c}+4 a^3 \mu^\text{c}},
\ee
and the correct equilibrium solution~\eqref{eq:Displacements1} is contained herein and attained when choosing
\be\label{eq:UaNeumannCorr}
	u_a = \frac{a \, b^3 p\, (3 \kappa^\text{c}+4 \mu^\text{c})}
		{12 a^3 (\kappa^\text{i}-\kappa^\text{c}) \mu^\text{c}+3 b^3 \kappa^\text{c} (3 \kappa^\text{i}+4 \mu^\text{c})}.
\ee
The total potential energy in this case is
\be
\begin{split}
	\tilde{W} & = \frac{1}{2}\int_\Omega\tilde{\bfepsilon}\cdot\dsC\,\tilde{\bfepsilon}\,\dd v
			- \int_{\partial\Omega} p\,\bfn\cdot\bfu \, \dd s\\
		& = \frac{1}{2}\int_0^a \left[\kappa^\text{i}(\tr\tilde{\bfepsilon}^\text{i})^2 
			+ \mu^\text{i}\,\tilde{\bfepsilon}_\text{dev}^\text{i}\cdot\tilde{\bfepsilon}_\text{dev}^\text{i}\right]\,4\pi r^2 \,\dd r
			+ \frac{1}{2}\int_a^b \left[\kappa^\text{c}(\tr\tilde{\bfepsilon}^\text{c})^2 
			+ \mu^\text{c}\,\tilde{\bfepsilon}_\text{dev}^\text{c}\cdot\tilde{\bfepsilon}_\text{dev}^\text{c}\right]\,4\pi r^2 \,\dd r
			- 4\pi b^2\,p\,u_r^\text{c}(b)\\
		& = \frac{4 \,a \,\pi\,  u_a}{3 b^3 \kappa^\text{c}+4 a^3 \mu^\text{c}}
		  \left[12 a^3 u_a  (\kappa^\text{i}-\kappa^\text{c}) \mu^\text{c}+3 b^3 u_a  \kappa^\text{c} 
		  (3 \kappa^\text{i}+4 \mu^\text{c})-a b^3 p (3 \kappa^\text{c}+4 \mu^\text{c})\right].
\end{split}
\ee
By applying the variational principle~\eqref{eq:DirichletCondEigen}, the equilibrium solution can be obtained by minimization, viz.
\be
	\tilde{u}_a = \argmin \tilde{W} \qquad\Rightarrow\qquad
	\frac{\partial \tilde{W}}{\partial u_a} = 0,
\ee
which yields the correct equilibrium solution~\eqref{eq:UaNeumannCorr}. To signal whether this equilibrium solution corresponds to an energy minimum, we compute
\be
	\frac{\partial^2 \tilde{W}}{\partial u_a^2}
	= \frac{24 a \pi  \left(4 a^3 (\kappa^\text{i}-\kappa^\text{c}) \mu^\text{c}+b^3 \kappa^\text{c} 
		(3 \kappa^\text{i}+4 \mu^\text{c})\right)}{3 b^3 \kappa^\text{c}+4 a^3 \mu^\text{c}}
	\begin{cases}
	> 0, & \ \text{if} \quad \kappa^\text{i} > 
		-\cfrac{4 \left(b^3-a^3\right)\kappa^\text{c} \mu^\text{c}}{3 b^3 \kappa^\text{c}+4 a^3 \mu^\text{c}} = \kappa^\text{i}_0, \\
	\leq 0, & \ \text{if}\quad \kappa^\text{i} \leq 
		-\cfrac{4 \left(b^3-a^3\right)\kappa^\text{c} \mu^\text{c}}{3 b^3 \kappa^\text{c}+4 a^3 \mu^\text{c}} = \kappa^\text{i}_0.
	\end{cases}
\ee
Therefore, the equilibrium solution may correspond to a minimum of the total potential energy until the effective bulk modulus first reaches $0$ with decreasing inclusion bulk modulus $\kappa^\text{i}$; i.e.\ the Neumann boundary value problem imposes stronger restrictions on the inclusion elastic moduli than the Dirichlet problem. We carefully say that the solution \emph{may} correspond to a minimum because we only checked for a fairly-limited space of perturbations, while a more general analysis confirms the found solution is indeed a minimizer.

\subsection{Neumann boundary value problem: inapplicability of the dual variational principle}
\label{sec:dual}

We can use the same example to prove that the dual variational principle does not hold, i.e.\ in case of non-positive-definite elastic moduli in all or part of a statically stable body the equilibrium solution is \emph{not} in general the minimizer of
\be\label{eq:dualPrinciple}
	\overline{W} = \inf_{\bfsigma(\bfx) \atop \bfsigma(\bfx)\bfn(\bfx)=\bft(\bfx) \ \text{on} \ \partial\Omega}
	\int_\Omega \frac{1}{2}\,\bfsigma(\bfx)\cdot \dsC^{-1}(\bfx) \,\bfsigma(\bfx) \,\dd V.
\ee
For an isotropic, linear elastic solid Hooke's law
\be
	\bfepsilon = \dsC^{-1}\bfsigma = \frac{1}{2\mu}\left(\bfsigma - \frac{3\kappa-2\mu}{9\kappa}(\tr\bfsigma)\,\id\right)
\ee
reduces the energy density in dual form to
\be\label{eq:IsotropicStressInequality}
	\left\langle \bfsigma(\bfx)\cdot\dsC^{-1}(\bfx)\,\bfsigma(\bfx)\right\rangle
	= \left\langle \frac{1}{2\mu(\bfx)} \bfsigma(\bfx)\cdot\bfsigma(\bfx)
		- \frac{3\kappa(\bfx)-2\mu(\bfx)}{9\kappa(\bfx)} \left[\tr\bfsigma(\bfx)\right]^2\right\rangle.
\ee
Again, we seek solutions of the Lam\'{e} type with stresses
\begin{subequations}
\begin{align}
	\bfsigma^\text{i}(r) & = \sigma_{rr}^\text{i}(r)\,\bfe_r\otimes\bfe_r 
		+ \sigma_{\varphi\varphi}^\text{i}(r)\,(\bfe_\varphi\otimes\bfe_\varphi + \bfe_\theta\otimes\bfe_\theta), 
		\qquad \sigma_{rr}^\text{i}(r) = A, \quad \sigma_{\varphi\varphi}^\text{i}(r) = A\\
	\bfsigma^\text{c}(r) & = \sigma_{rr}^\text{c}(r)\,\bfe_r\otimes\bfe_r 
		+ \sigma_{\varphi\varphi}^\text{c}(r)\,(\bfe_\varphi\otimes\bfe_\varphi + \bfe_\theta\otimes\bfe_\theta), 
		\qquad \sigma_{rr}^\text{i}(r) = B + \frac{C}{r^3}, \quad \sigma_{\varphi\varphi}^\text{i}(r) = B - \frac{C}{r^3}.
\end{align}\label{eq:Stresses}%
\end{subequations}
The boundary and continuity conditions are again~\eqref{eq:NeumannBCs}, which results in the same solution given by~\eqref{eq:Displacements} with constants~\eqref{eq:NeumannSolution}.


Let us verify whether or not this stable equilibrium solution can be found by application of the dual variational principle~\eqref{eq:dualPrinciple}. By analogy with the previous cases, let us construct a space of stress fields that are continuous inside the solid and satisfy the traction boundary condition, such that the space of solutions contains the equilibrium solution. Specifically, consider stresses $\overline{\bfsigma}(r)$ identical to~\eqref{eq:Stresses} whose coefficients $\overline{A}$, $\overline{B}$ and $\overline{C}$ are determined by enforcing 
\be
	\overline{\sigma}_{rr}^\text{c}(b) = p, \qquad
	\overline{\sigma}_{rr}^\text{i}(a) = \sigma_a, \qquad 
	\overline{\sigma}_{rr}^\text{c}(a) = \sigma_a
\ee	
with some radial interface traction $\sigma_a$. This results in
\be
	\overline{A} = \sigma_a,\quad 
	\overline{B} = \frac{a^3 \sigma_a - b^3 p}{b^3-a^3},\quad
	\overline{C} = \frac{a^3 b^3 (\sigma_a-p)}{b^3-a^3},
\ee
which is identical to the equilibrium solution~\eqref{eq:Displacements} with constants~\eqref{eq:NeumannSolution} if
\be\label{eq:sigmaa}
	\sigma_a = \frac{b^3 p \kappa^\text{i} (3 \kappa^\text{c}+4 \mu^\text{c})}{4 a^3 (\kappa^\text{i}-\kappa^\text{c}) \mu^\text{c}+b^3 \kappa^\text{c} (3 \kappa^\text{i}+4 \mu^\text{c})}.
\ee
For this specific space of stress distributions, the dual total energy of the two-phase solid can be written as
\be
\begin{split}
	\overline{W} & = \frac{1}{2}\int_\Omega\overline{\bfsigma}\cdot\dsC^{-1}\,\overline{\bfsigma}\,\dd V
		= \frac{1}{2}\int_\Omega\left[\frac{1}{2\mu} \overline{\bfsigma}\cdot\overline{\bfsigma}
		- \frac{3\kappa-2\mu}{9\kappa} (\tr \overline{\bfsigma})^2\right]\,\dd V\\
	& = \frac{1}{2}\int_0^a\left[\frac{1}{2\mu^\text{i}} \overline{\bfsigma}^\text{i}\cdot\overline{\bfsigma}^\text{i}
		- \frac{3\kappa^\text{i}-2\mu^\text{i}}{9\kappa^\text{i}} (\tr \overline{\bfsigma}^\text{i})^2\right]\,4\pi r^2 \dd r + 
		\frac{1}{2}\int_a^b\left[\frac{1}{2\mu^\text{c}} \overline{\bfsigma}^\text{c}\cdot\overline{\bfsigma}^\text{c}
		- \frac{3\kappa^\text{c}-2\mu^\text{c}}{9\kappa^\text{c}} (\tr \overline{\bfsigma}^\text{c})^2\right]\,4\pi r^2 \dd r\\
	& = \frac{\pi  \left[4 b^6 p^2 \kappa^\text{i} \mu^\text{c}+4 a^6 (\kappa^\text{i}-\kappa^\text{c}) \mu^\text{c} \sigma_a^2+a^3 b^3 \left(3 p^2 \kappa^\text{i} \kappa^\text{c}-2 p \kappa^\text{i} (3 \kappa^\text{c}+4 \mu^\text{c}) \sigma_a+\kappa^\text{c} (3 \kappa^\text{i}+4 \mu^\text{c}) \sigma_a^2\right)\right]}{3 \left(b^3-a^3\right) \kappa^\text{i} \kappa^\text{c} \mu^\text{c}}.
\end{split}
\ee
Equilibrium solutions are sought by identifying stress fields which render the total dual potential energy stationary. Application of
\be
	\frac{\partial \overline{W}}{\partial \sigma_a} = 0
\ee
indeed yields the correct equilibrium solution~\eqref{eq:sigmaa}. However, it is a simple exercise to show that
\be
	\frac{\partial^2 \overline{W}}{\partial \sigma_a^2} 
	= -\frac{2 a^3 \pi  \left[4 a^3 (\kappa^\text{i}-\kappa^\text{c}) \mu^\text{c}+b^3 \kappa^\text{c} (3 \kappa^\text{i}+4 \mu^\text{c})\right]}{3 \left(a^3-b^3\right) \kappa^\text{i} \kappa^\text{c} \mu^\text{c}}
	\begin{cases}
	< 0, \quad & \text{if} \ 0 > \kappa_1 > 
		-\cfrac{4 \left(b^3-a^3\right)\kappa^\text{c} \mu^\text{c}}{3 b^3 \kappa^\text{c}+4 a^3 \mu^\text{c}}
		= \kappa^\text{i}_0,\\
	\geq 0, \quad & \text{if} \ \kappa_1 \geq 0 \quad \text{or} \quad
	\kappa_1 \leq -\cfrac{4 \left(b^3-a^3\right)\kappa^\text{c} \mu^\text{c}}{3 b^3 \kappa^\text{c}+4 a^3 \mu^\text{c}}
		= \kappa^\text{i}_0.
	\end{cases}
\ee
This implies that, when using the dual variational principle, the equilibrium solution can still be found from stationarity of the total potential energy. When considering an overall positive-definite inhomogeneous body ($\kappa^\text{i},\kappa^\text{c}>0$), the variational principle~\eqref{eq:dualPrinciple} still applies and the stable equilibrium solution corresponds to a global energy minimum. However, as soon as the inclusion phase violates positive-definiteness (when $\kappa^\text{i}<0$), the variational principle~\eqref{eq:dualPrinciple} no longer applies: the regime $0 > \kappa_1 > \kappa^\text{i}_0$, which was shown before to possess a stable equilibrium solution corresponding to a global energy minimum, now turns into an energy maximum. Likewise, $\kappa_1 \leq \kappa^\text{i}_0$ is signaled to possess an equilibrium solution which minimizes the dual potential energy (to confirm, more general perturbations are required). Yet, we showed before that the Neumann problem for $\kappa_1 \leq \kappa^\text{i}_0$ does not possess any stable equilibrium solutions. In summary, the variational principle~\eqref{eq:dualPrinciple} does not apply and leads to incorrect stability conclusions if negative-stiffness phases are being considered.

\begin{figure}[tb]
	\centering
		\includegraphics[width=0.55\textwidth]{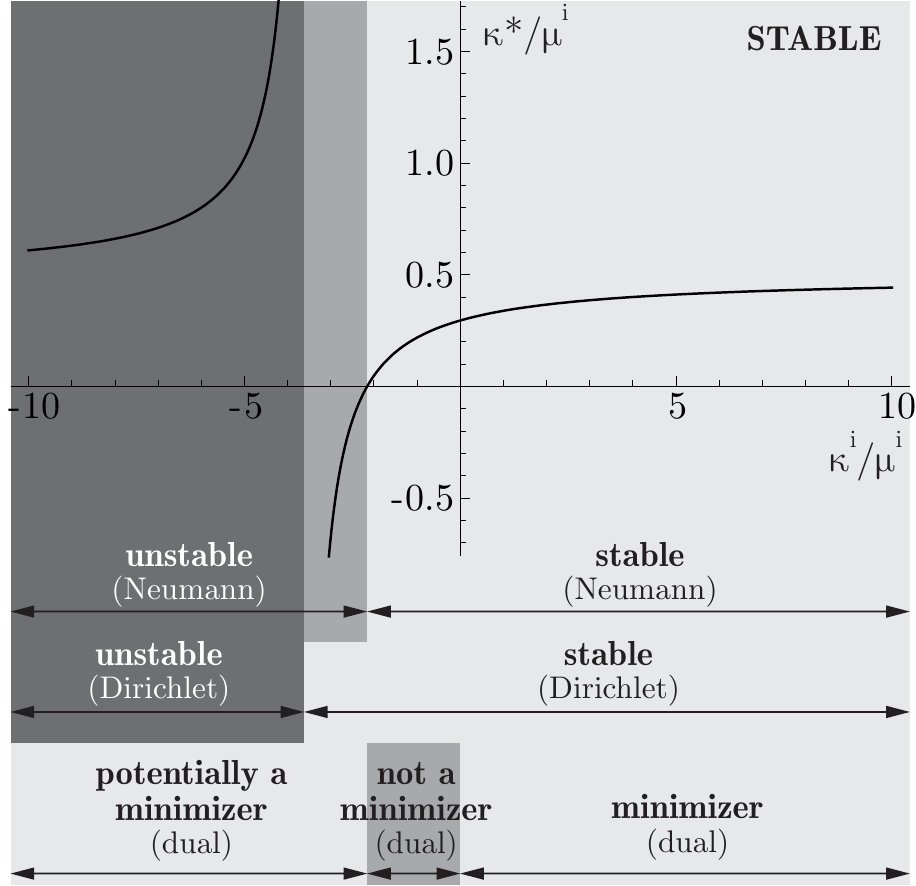}
	\caption{Effective bulk modulus of a coated sphere along with the stability conditions for the Dirichlet and Neumann problem, as well as stability conditions obtained using the dual variational principle. All results are for the example values of $\mu^\text{c}/\mu^\text{i}=2$, $a/b=1/2$, $\kappa^\text{c}/\mu^\text{c}=2$.}
	\label{fig:NeumannDirichlet}
\end{figure}

Figure~\ref{fig:NeumannDirichlet} illustrates the stability conditions obtained for both the Dirichlet and Neumann boundary value problems along with the incorrect stability conditions obtained by using the dual variational principle. Shown is the normalized effective bulk modulus~\eqref{eq:EffBulkSphere} vs.\ the normalized inclusion bulk modulus $\kappa^\text{i}$ for specific values of the remaining moduli.


\section{Effective elasticities of inhomogeneous bodies and composites}
\label{sec:EffectiveModuli}

The main focus of this contribution is on the effective elastic moduli of linear elastic inhomogeneous bodies and composites. Recently, it was shown that embedding phases with non-positive-definite elastic moduli (so-called \emph{negative stiffness}) in a composite has the potential to result in extreme effective composite properties including unbounded stiffness in elastic composites~\citep{LakesDrugan2002} as well as extremely high stiffness and damping in viscoelastic solids~\citep{Lakes2001a,Lakes2001b}. Non-positive-definite moduli raise questions of stability: as shown in the previous section, homogeneous elastic solids with pure traction boundary conditions cannot have a negative bulk modulus and be stable. In composites as well as in heterogeneous bodies, the geometric constraints among the various phases provide stabilization, which was shown to indeed permit negative-stiffness phases when embedded in a stiff matrix or coating~\citep{Drugan2007,KochmannDrugan2009,Kochmann2012,KochmannDrugan2012}. Simple structural examples have demonstrated that the amount of negative stiffness thus stabilized is insufficient to create extreme effective moduli~\citep{WojnarKochmann2013a,WojnarKochmann2013b}. Yet, to date no rigorous analysis has shown whether or not general linear elastic composite materials (of arbitrary geometry, phase arrangement, anisotropy, and constituent properties) can lead to extreme effective stiffness due to a negative-stiffness phase (or negative-stiffness phases). Therefore, in the following we will use those relations derived above to link stability conditions to the effective elastic moduli of linear elastic inhomogeneous bodies and composites.

\subsection{Voigt bounds}
\label{sec:Voigt}

Let us assume we have a solution $\tilde{\bfu}(\bfx)$ to the elasticity equation with affine boundary conditions $\tilde{\bfu}(\bfx)=\bfepsilon_0\bfx$ on $\partial\Omega$ with constant symmetric $\bfepsilon_0=\langle\tilde{\bfepsilon}(\bfx)\rangle$. Following Hill, we must obey energy equivalence of the form $\langle \tilde{\bfepsilon} \cdot\tilde{\bfsigma}\rangle = \langle \tilde{\bfepsilon} \rangle \cdot \langle\tilde{\bfsigma}\rangle$. We define the effective elasticity tensor for the case of Dirichlet boundary conditions by $\langle\tilde{\bfsigma}\rangle = \dsC_*^\text{D} \,\langle \tilde{\bfepsilon}\rangle$. Therefore, if $\tilde{\bfu}(\bfx)$ is stable, the variational principle~\eqref{eq:infcondition} applies, i.e.\ we have:
\be\label{eq:Minimizer}
	\bfepsilon_0\cdot\dsC_*^\text{D}\,\bfepsilon_0= 
	\langle \tilde{\bfepsilon} \rangle \cdot \langle\tilde{\bfsigma}\rangle =
	\langle \tilde{\bfepsilon} \cdot\tilde{\bfsigma}\rangle = 
	\inf_{\Large \bfu(\bfx) \atop \bfu(\bfx)=\bfepsilon_0\bfx \ \text{on} \ \partial\Omega}
		\langle \bfepsilon(\bfx)\cdot \dsC(\bfx) \,\bfepsilon(\bfx) \rangle.
\ee
For the particular case of $\bfepsilon(\bfx)=\bfepsilon_0 = \text{const.}$, we see that
\be \label{eq:Voigtbound}
	\dsC_*^\text{D} \leq \langle \dsC(\bfx) \rangle,
\ee
which corresponds to the classical Voigt bound. The tensor inequality $\dsC_1 \leq \dsC_2$ implies $\bfT\cdot\dsC_1\,\bfT \leq \bfT\cdot\dsC_2\,\bfT$ for all real second-order tensors $\bfT$. The dual variational principle does not apply, as shown Section~\ref{sec:CoatedSphere}, and so it is not immediately clear if the Reuss bound holds or not. Note that our goal here was not rederive the classical bounds; instead we have shown that, \emph{if an elastic inhomogeneous body is in stable equilibrium, then stability requires the effective moduli to satisfy~\eqref{eq:Voigtbound}}, which severely restricts the space of attainable effective elastic moduli.

In case of Neumann boundary conditions, let us assume a solution $\tilde{\bfu}(\bfx)$ to the elasticity equation with $\tilde{\bft}(\bfx)=\bfsigma_0\bfn$ on $\partial\Omega$ with constant symmetric $\bfsigma_0=\langle\tilde{\bfsigma}(\bfx)\rangle$, and we define the effective compliance tensor for the Neumann problem via $\langle\tilde{\bfepsilon}\rangle = \dsS_*^\text{N} \,\langle \tilde{\bfsigma}\rangle$. If $\tilde{\bfu}(\bfx)$ is stable, the variational principle~\eqref{eq:NeumannVariationalP} holds and therefore we have
\be\label{eq:NeumannTrial}
\begin{split}
	\int_{\Omega} \frac{1}{2} \tilde{\bfepsilon}(\bfx)\cdot \dsC(\bfx) \, \tilde{\bfepsilon}(\bfx)\,\dd V
		- \int_{\partial\Omega} \bft_0(\bfx)\cdot \tilde{\bfu}(\bfx)\,\dd S 
	= \inf_{\bfu(\bfx)} \left\{\int_{\Omega} \frac{1}{2} \bfepsilon(\bfx)\cdot \dsC(\bfx) \, \bfepsilon(\bfx)\,\dd V
		- \int_{\partial\Omega} \bft_0(\bfx)\cdot \bfu(\bfx)\,\dd S \right\}
\end{split}
\ee
for any admissible strain field $\bfepsilon(\bfx)$. Using $\bft_0(\bfx)=\bfsigma_0\bfn(\bfx)$ we obtain
\be
\begin{split}
	\langle\tilde{\bfepsilon}\rangle = \dsS_*^\text{N}\,\langle\tilde{\bfsigma}\rangle
	 & = \dsS_*^\text{N}\,\frac{1}{V}\int_\Omega \tilde{\bfsigma} \,\dd V
	 = \dsS_*^\text{N}\,\frac{1}{V}\int_{\partial\Omega} \sym\left(\tilde{\bft}\otimes\bfx\right) \,\dd S
	 = \dsS_*^\text{N}\,\frac{1}{V}\int_{\partial\Omega} \sym\left(\bft_0\otimes\bfx\right) \,\dd S\\
	 & = \dsS_*^\text{N}\,\frac{1}{V}\int_{\Omega} \bfsigma_0\,\dd V
	 = \frac{\dsS_*^\text{N}\,\bfsigma_0}{V} 
	 = \dsS_*^\text{N}\,\langle\bfsigma_0\rangle .
\end{split}
\ee
Consequently, the left-hand side of~\eqref{eq:NeumannTrial} becomes (applying the divergence theorem and using $\divv\tilde{\bfsigma}=\bfnull$)
\be
\begin{split}
	\int_{\Omega} \frac{1}{2} \tilde{\bfepsilon}(\bfx)\cdot \tilde{\bfsigma}(\bfx)\,\dd V
		- \int_{\partial\Omega} \bft_0(\bfx)\cdot \tilde{\bfu}(\bfx)\,\dd S 
	& = \frac{1}{2} \int_{\partial\Omega} \bft_0(\bfx)\cdot \tilde{\bfu}(\bfx)\,\dd V
		- \int_{\partial\Omega} \bft_0(\bfx)\cdot \tilde{\bfu}(\bfx)\,\dd S \\
	& = - \frac{1}{2} \bfsigma_0\cdot\int_\Omega\tilde{\bfepsilon}(\bfx)\,\dd V
	= - \frac{V}{2} \bfsigma_0\cdot\dsS_*^\text{N}\,\bfsigma_0.
\end{split}
\ee
We can find a rigorous upper bound to the right-hand side of~\eqref{eq:NeumannTrial} as follows. Assume a constant trial strain field $\bfepsilon(\bfx)=\check{\bfepsilon}$ so that
\be\label{eq:ConstantTrialField}
	\inf_{\bfu(\bfx)} \left\{\int_{\Omega} \frac{1}{2} \bfepsilon(\bfx)\cdot \dsC(\bfx) \, \bfepsilon(\bfx)\,\dd V
		- \int_{\partial\Omega} \bft_0(\bfx)\cdot \bfu(\bfx)\,\dd S \right\}
	\leq V\cdot\inf_{\large\check{\bfepsilon}} \left\{
	\frac{1}{2}\check{\bfepsilon}\cdot \langle\dsC(\bfx)\rangle\,\check{\bfepsilon} - \bfsigma_0\cdot \check{\bfepsilon}\right\}
\ee
and minimization with respect to $\check{\bfepsilon}$ yields
\be
	\check{\bfepsilon} = \langle\dsC(\bfx)\rangle^{-1}\bfsigma_0 \qquad \Rightarrow \qquad
	\inf_{\large\check{\bfepsilon}} \left\{
	\frac{1}{2}\check{\bfepsilon}\cdot \langle\dsC(\bfx)\rangle\,\check{\bfepsilon} - \bfsigma_0\cdot \check{\bfepsilon}\right\} 
	= -\frac{1}{2}\bfsigma_0\cdot\langle\dsC(\bfx)\rangle^{-1}\bfsigma_0.
\ee
Note that here we assume $\langle\dsC(\bfx)\rangle$ is positive-definite, which, however, is a stability requirement. Specifically, we showed that if the solid is to be statically stable, the variational principle~\eqref{eq:NeumannTrial} applies. However, existence of a minimizer of~\eqref{eq:NeumannTrial} (with bounded strains) requires that $\langle\dsC(\bfx)\rangle$ is positive-definite because of~\eqref{eq:ConstantTrialField}. Altogether, this results in
\be
	- \frac{1}{2} \bfsigma_0\cdot\dsS_*^\text{N}\,\bfsigma_0 \leq -\frac{1}{2}\bfsigma_0\cdot\langle\dsC\rangle^{-1}\bfsigma_0
	\qquad \Leftrightarrow \qquad
	\bfsigma_0\cdot\dsS_*^\text{N}\,\bfsigma_0 \geq \bfsigma_0\cdot\langle\dsC(\bfx)\rangle^{-1}\bfsigma_0
\ee
which implies $\dsS_*^\text{N}$ is positive definite, and so introducing $\dsC_*^\text{N}=\left(\dsS_*^\text{N}\right)^{-1}$ we have finally
\be
	0\leq\dsC_*^\text{N} \leq \langle\dsC(\bfx)\rangle,
\ee
which again enforces the classical Voigt bound and is the analogue of~\eqref{eq:Voigtbound} for the Neumann boundary value problem. This proves that, \emph{if a heterogeneous elastic solid in static equilibrium is to be stable, the effective modulus tensor obtained from both the Neumann and Dirichlet boundary value problems must be such that $\dsC_*\leq\langle\dsC\rangle$}, which specifically excludes the stability of extreme effective moduli surpassing those of the individual constituents of an inhomogeneous body. In the following, we simply refer to the effective moduli, implying that our conclusions hold for effective moduli obtained from both Neumann and Dirichlet conditions.

The effective bulk modulus $\kappa_*$ of an isotropic medium is linked to the effective modulus tensor via
\be
	\kappa_* = \frac{1}{9}\,\id\cdot\dsC_*\,\id.
\ee
Consequently, stability restricts the effective bulk modulus of a heterogeneous elastic solid (with either affine displacement or the aforementioned traction boundary conditions) to
\be\label{eq:VoigtKappa}
	\kappa_* \leq \langle \kappa(\bfx)\rangle,
\ee 
which excludes the stability of an infinite bulk modulus unless one constituent possesses infinite bulk stiffness (is incompressible). Note that for Neumann boundary conditions we have established $\dsS_*^\text{N}\geq 0$ for stability which further implies $\kappa_*^\text{N}\geq 0$.


Similarly, the effective shear modulus $\mu_*$ can be expressed as
\be
	\mu_* = \frac{1}{2}\bfe\cdot\dsC_*\bfe \qquad \text{for any symmetric tensor}\ \bfe \quad \text{with} \quad \tr\bfe=0 \quad\text{and}
	\quad \bfe\cdot\bfe=1.
\ee
Therefore, we arrive at the analogous condition
\be\label{eq:VoigtMu}
	\mu_*\leq \langle \mu(\bfx)\rangle,
\ee
which rules out the stability of extreme values of the effective shear modulus. For Neumann boundary conditions we again have $\dsS_*^\text{N}\geq 0$ which implies the stability requirement $\mu_*^\text{N}\geq 0$.


To further rule out inhomogeneous bodies with static infinite stiffness due to a negative-stiffness phase, we can compare the effective moduli of two heterogeneous solids which differ by their local elastic moduli. Let us reduce the elastic moduli from $\dsC(\bfx)$ to $\dsC'(\bfx)\leq\dsC(\bfx)$ in all or some subpart of body $\Omega$ so that
\be
	\langle\bfepsilon\rangle \cdot(\dsC_*')^\text{D}\langle\bfepsilon\rangle 
	= \inf_{\Large \bfu(\bfx) \atop \bfu(\bfx)=\bfv(\bfx) \ \text{on} \ \partial\Omega}
		\langle \bfepsilon(\bfx)\cdot \dsC'(\bfx) \,\bfepsilon(\bfx) \rangle
	\leq
		\inf_{\Large \bfu(\bfx) \atop \bfu(\bfx)=\bfv(\bfx) \ \text{on} \ \partial\Omega}
		\langle \bfepsilon(\bfx)\cdot \dsC(\bfx) \,\bfepsilon(\bfx) \rangle 
	= \langle\bfepsilon\rangle \cdot\dsC_*^\text{D}\langle\bfepsilon\rangle,
\ee
and the analogous inequality holds for the Neumann problem, which implies that
\be
	\dsC'(\bfx)\leq\dsC(\bfx) \quad \forall \ \bfx\in\Omega \qquad \Rightarrow \qquad \dsC_*' \leq \dsC_*
	\quad \Leftrightarrow \quad 
	\kappa_*' \leq \kappa_*, \quad \mu_*'\leq\mu_*.
\ee
In other words, if we reduce the bulk (shear) modulus of one of the phases in an inhomogeneous body, then the effective bulk (shear) modulus must also decrease when assuming stability, i.e.\ it must be a monotonic function of the moduli of each phase.

In conclusion, we have shown that in linear elastic inhomogeneous bodies (i) an \emph{infinite effective bulk or shear modulus is always unstable} unless one of the constituents has infinite stiffness, and (ii) reducing the bulk modulus of one of the phases from positive to negative values cannot lead to an increase in the effective bulk modulus if the inhomogeneous body is overall stable. This confirms results obtained for specific two-phase solids and composites~\citep{KochmannDrugan2012,WojnarKochmann2013a,WojnarKochmann2013b} and greatly generalizes those findings to arbitrary elastic inhomogeneous bodies. Note that our analysis holds for arbitrary phase arrangement and arbitrarily many phases having arbitrary (strongly-elliptic elastic) moduli.

The variational principle~\eqref{eq:NeumannVariationalP} also demonstrates that an inhomogeneous solid whose phases are compressible cannot exhibit an effective infinite bulk stiffness $\kappa_*^\text{N}$ under Neumann boundary conditions. To show this, we can surround the inhomogeneous body by a shell of compressible fluid with bulk modulus $\kappa_0$. Then the variational principle~\eqref{eq:infcondition} applies and we see that $\tilde{\bfu}(\bfx)$ must minimize
\be
	W = \inf_{\Large\bfu(\bfx) \atop \bfu(\bfx)=\bfv(\bfx) \ \text{on} \ \partial\Omega} \frac{1}{2}
		\left\{\int_{\text{\scriptsize shell}} \kappa_0 \left[\tr\bfepsilon(\bfx)\right]^2 \dd V + 
		\int_{\text{composite}} \bfepsilon(\bfx)\cdot \dsC(\bfx) \,\bfepsilon(\bfx) \,\dd V\right\}.
\ee
When we apply radial displacements $\bfu(\bfx)=\alpha\,\bfn$ on the outside boundary with outward unit normal $\bfn$ and constant $\alpha$, an infinite effective bulk modulus $\kappa_*^\text{N}$ in the inhomogeneous body will result in it not changing its volume. Therefore, if we let the surrounding shell be very thin, the pressure and thus the energy inside the thin shell will be enormous. However, when considering the exact heterogenous moduli inside the inhomogeneous body, we can alternatively choose a constant field $\bfepsilon(\bfx)=\bfepsilon$ inside the body and the total energy will be much lower. So, $\tilde{\bfu}(\bfx)$ is unstable. In other words, a heterogeneous body having a non-infinite bulk modulus everywhere inside the solid cannot result in an infinite static effective bulk modulus under Neumann conditions.


\subsection{Hashin-Shtrikman variational principles}
\label{sec:HS}

In the remainder of the paper we assume the body contains a periodic or statistically homogeneous composite with microstructure much smaller than the dimensions of $\Omega$, so that $\dsC_*^\text{N}=\dsC_*^\text{D}=\dsC_*$, where $\dsC_*$ is the effective elasticity tensor of the composite. However, some of the results extend to inhomogeneous bodies using the arguments of~\citet{Milton2012}. Before deriving tighter bounds for isotropic two-phase composites by taking specific strain trial fields, let us review the Hashin-Shtrikman variational inequalities as to their validity in the presence of locally non-positive-definite elastic moduli. For the lower Hashin-Shtrikman variational principle this was done already [see, for example, \mbox{Section 13.5} of~\citep{Milton2002}] but for completeness we include that treatment. \citet{HashinShtrikman1963} introduced a reference medium with constant elastic modulus tensor $\dsC_0$ as well as a polarization field
\be\label{eq:polarization}
	\bfp(\bfx) = \left[\dsC(\bfx)-\dsC_0\right]\bfepsilon(\bfx) = \bfsigma(\bfx) - \dsC_0\,\bfepsilon(\bfx).
\ee
For any trial polarization field $\bfp(\bfx)$, we introduce the operator $\bfGamma(\bfx)$ such that $\bfGamma(\bfx)\bfp(\bfx)$ is a strain with $\langle\bfGamma(\bfx)\bfp(\bfx)\rangle=0$ and $\bfp(\bfx)-\dsC_0\bfGamma(\bfx) \bfp(\bfx)$ is a stress, i.e.\ $\divv(\bfp-\dsC_0\bfGamma \bfp)=\bfnull$. When $\bfp(\bfx)$ is the actual polarization field then~\eqref{eq:polarization} implies
\be
	\bfGamma(\bfx)\,\bfp(\bfx) = \langle\bfepsilon\rangle - \bfepsilon(\bfx)
\ee
and the volume average of~\eqref{eq:polarization} gives
\be\label{eq:HSeq1}
	\left[\id + \left(\dsC(\bfx)-\dsC_0\right)\bfGamma(\bfx)\right]\bfp(\bfx) = \left[\dsC(\bfx)-\dsC_0\right]\langle\bfepsilon\rangle.
\ee
$\bfGamma$ is self-adjoint and satisfies $\bfGamma=\bfGamma \,\dsC_0 \bfGamma$, and therefore if $\dsC_0$ is positive-semi-definite for strains (i.e.\ if $\dsC_0$ is elliptic), and since $\bfGamma\bfp$ is a strain, it follows that $\bfGamma$ is positive-semi-definite. We choose the reference medium such that
\be\label{eq:Choices}
	\dsC(\bfx) > \dsC_0
\ee
everywhere. Therefore, $\dsC(\bfx)-\dsC_0$ is positive-definite and can be inverted, which turns~\eqref{eq:HSeq1} into
\be\label{eq:HS1}
	\left[\left(\dsC(\bfx)-\dsC_0\right)^{-1} + \bfGamma(\bfx)\right]\bfp(\bfx) = \langle\bfepsilon\rangle.
\ee
Taking volume averages in~\eqref{eq:polarization} gives
\be\label{eq:HS2}
	\langle\bfp\rangle = \langle\bfsigma\rangle - \dsC_0\langle\bfepsilon\rangle
	= \left(\dsC_*-\dsC_0\right)\langle\bfepsilon\rangle.
\ee
Note that~\eqref{eq:Choices} along with the properties of $\bfGamma$ also guarantees that $(\dsC-\dsC_0)^{-1} + \bfGamma$ is positive definite so that
\be\label{eq:HSVarPr}
	\left\langle 
	\left[\bfp(\bfx)-\langle\bfp\rangle\right]\cdot\left[(\dsC(\bfx)-\dsC_0)^{-1} + \bfGamma(\bfx)\right]\left[\bfp(\bfx)-\langle\bfp\rangle\right]
	\right\rangle\geq 0.
\ee
By using~\eqref{eq:HS1} and \eqref{eq:HS2}, we expand~\eqref{eq:HSVarPr} into the Hashin-Shtrikman variational inequality
\be\label{eq:HSLowerInequality}
	\langle\bfepsilon\rangle \cdot\dsC_* \,\langle\bfepsilon\rangle\geq 
		\langle\bfepsilon\rangle\cdot\dsC_0\,\langle\bfepsilon\rangle + 2\,\langle\bfepsilon\rangle\cdot \langle\bfp(\bfx)\rangle
		-\left\langle \bfp(\bfx)\cdot\left[(\dsC(\bfx)-\dsC_0)^{-1} + \bfGamma(\bfx)\right]\bfp(\bfx)\right\rangle,
\ee
which holds for all admissible choices of the average strain field $\langle\bfepsilon\rangle$ and of the trial polarization field $\bfp(\bfx)$, and which yields the Hashin-Shtrikman lower bound for isotropic composites. For the optimal choice 
\be
	\langle\bfepsilon\rangle = (\dsC_*-\dsC_0)^{-1}\langle\bfp\rangle,
\ee
the variational inequality can be rewritten as a variational principle:
\be
	\bfp_0 \cdot (\dsC_*-\dsC_0)^{-1}\bfp_0
	= \inf_{\Large \bfp(\bfx) \atop \langle\bfp(\bfx)\rangle = \bfp_0} \left\langle\bfp(\bfx)\cdot 
	\left[(\dsC(\bfx)-\dsC_0)^{-1} + \bfGamma\right]\bfp(\bfx)\right\rangle.
\ee

Next, following the approach of~\citet{Hill1963b} who derived the Hashin-Shtrikman variational principles from the classical variational principles, consider a reference medium such that
\be\label{eq:HSOtherCase}
	\dsC_0 > \dsC(\bfx)
\ee
with $\dsC_0$ being positive-definite. The variational principle~\eqref{eq:infcondition} can be recast into the variational inequality
\be\label{eq:VarIneqHill}
	\langle\tilde{\bfepsilon}\rangle \cdot \dsC_* \,\langle\tilde{\bfepsilon}\rangle
	\leq
	\langle\bfepsilon(\bfx)\cdot \dsC_0 \,\bfepsilon(\bfx)\rangle
	- \left\langle\bfepsilon(\bfx) \cdot \left[\dsC(\bfx) - \dsC_0\right] \,\bfepsilon(\bfx)\right\rangle
\ee
for any trial field $\bfepsilon(\bfx)$. Because of~\eqref{eq:HSOtherCase},  $\delta\dsC(\bfx)=\dsC(\bfx)-\dsC_0$ is negative-definite, which allows us to expand
\be
	\left\langle\bfepsilon(\bfx)\cdot \delta\dsC(\bfx)\,\bfepsilon(\bfx) \right\rangle
	= \left\langle\left(\bfp(\bfx)-\delta\dsC(\bfx)\,\bfepsilon(\bfx)\right)
		\cdot \delta\dsC^{-1}(\bfx)
		\left(\bfp(\bfx)-\delta\dsC(\bfx)\,\bfepsilon(\bfx)\right)
		\right\rangle
	+ 2 \left\langle\bfepsilon(\bfx)\cdot\bfp(\bfx)\right\rangle
	- \left\langle \bfp(\bfx)\cdot \delta\dsC^{-1}(\bfx)\bfp(\bfx)\right\rangle
\ee
for any polarization field $\bfp(\bfx)$. Insertion into~\eqref{eq:VarIneqHill} yields
\be\label{eq:HillsInequality0}
\begin{split}
	\langle\bfepsilon\rangle \cdot\dsC_* \,\langle\bfepsilon\rangle
		\leq \langle\bfepsilon(\bfx)\cdot\dsC_0\,\bfepsilon(\bfx)\rangle 
	& + 2\,\langle\bfepsilon(\bfx)\rangle\cdot \langle\bfp(\bfx)\rangle
		- \left\langle \bfp(\bfx)\cdot\delta\dsC^{-1}(\bfx)\,\bfp(\bfx)\right\rangle\\
	& + \left\langle \left[\bfp(\bfx)-\delta\dsC(\bfx)\,\bfepsilon(\bfx)\right]\cdot\delta\dsC^{-1}(\bfx)
			\left[\bfp(\bfx)-\delta\dsC(\bfx)\,\bfepsilon(\bfx)\right]\right\rangle.
\end{split}
\ee
We substitute the trial field 
\be
	\hat{\bfepsilon}(\bfx) = \langle\bfepsilon\rangle - \bfGamma(\bfx)\,\bfp(\bfx)
\ee
into the variational inequality~\eqref{eq:HillsInequality0} and use the facts that $\bfGamma\,\dsC_0\,\langle\bfepsilon\rangle=\bfnull$ and $\bfGamma\,\dsC\,\bfGamma=\bfGamma$, which results in the bound~\citep{Hill1963b}
\be\label{eq:HillsInequality}
\begin{split}
	\langle\bfepsilon\rangle \cdot\dsC_* \,\langle\bfepsilon\rangle
		\leq \langle\bfepsilon\rangle\cdot\dsC_0\,\langle\bfepsilon\rangle 
	& + 2\,\langle\bfepsilon\rangle\cdot \langle\bfp(\bfx)\rangle
		- \left\langle \bfp(\bfx)\cdot\left[\delta\dsC^{-1}(\bfx) + \bfGamma(\bfx)\right]\bfp(\bfx)\right\rangle \\
	& + \left\langle \left[\bfp(\bfx)-\delta\dsC(\bfx)\,\hat{\bfepsilon}(\bfx)\right]\cdot\delta\dsC^{-1}(\bfx)
			\left[\bfp(\bfx)-\delta\dsC(\bfx)\,\hat{\bfepsilon}(\bfx)\right]\right\rangle.
\end{split}
\ee
Except for the final term, this coincides with the other Hashin-Shtrikman variational inequality:
\be\label{eq:HSUpperInequality}
	\langle\bfepsilon\rangle \cdot\dsC_* \,\langle\bfepsilon\rangle
		\leq \langle\bfepsilon\rangle\cdot\dsC_0\,\langle\bfepsilon\rangle 
	+ 2\,\langle\bfepsilon\rangle\cdot \langle\bfp(\bfx)\rangle
		- \left\langle \bfp(\bfx)\cdot\left[(\dsC(\bfx)-\dsC_0)^{-1} + \bfGamma(\bfx)\right]\bfp(\bfx)\right\rangle.
\ee
However, notice that because of~\eqref{eq:HSOtherCase} the last term in~\eqref{eq:HillsInequality} is non-positive, so that the upper bound~\eqref{eq:HillsInequality} obtained from the classical variational principle is stronger than the Hashin-Shtrikman variational inequality~\eqref{eq:HSUpperInequality}. Because composite stability guarantees that the classical variational principle holds, the Hashin-Shtrikman variational inequality~\eqref{eq:HSUpperInequality} applies if the composite is stable, and hence so does the associated Hashin-Shtrikman upper bound for the elastic moduli of isotropic composites~\citep{HashinShtrikman1963}. In addition, if the trial polarization field equals the true polarization field then equality is obtained in~\eqref{eq:HSUpperInequality}, since the additional term in~\eqref{eq:HillsInequality} vanishes in this case. Therefore, the variational inequality again yields the associated Hashin-Shtrikman variational principle:
\be
	\bfp_0 \cdot (\dsC_0-\dsC_*)^{-1}\bfp_0
	= \inf_{\Large \bfp(\bfx) \atop \langle\bfp(\bfx)\rangle = \bfp_0} \left\langle\bfp(\bfx)\cdot 
	\left[(\dsC_0-\dsC(\bfx))^{-1} - \bfGamma\right]\bfp(\bfx)\right\rangle.
\ee

In summary, we have shown that, if the composite is statically stable, the classical Hashin-Shtrikman variational inequalities~\eqref{eq:HSLowerInequality} and~\eqref{eq:HSUpperInequality} apply, even if the composite phases violate positive-definiteness of their elasticities (as long as both phases are strongly-elliptic which is required for pointwise stability anyway). This also implies that if a well-ordered isotropic linear elastic composite is stable, \emph{the classical Hashin-Shtrikman upper and lower bounds apply and constrain the space of attainable effective elastic moduli of the composite}. We note that the remaining two Hashin-Shtrikman variational principles commonly obtained from the duality principle, see e.g.\ Section 13.5 of~\citep{Milton2002}, probably do not apply since we do not assume $\dsC(\bfx)$ is positive-definite everywhere.


\subsection{Upper bounds for isotropic two-phase composites}

Consider an isotropic two-phase composites with phase volume fractions $f_1$ and \mbox{$f_2=1-f_1$}. As shown in Section~\ref{sec:Voigt}, overall stability of the composite for both Dirichlet and Neumann boundary conditions enforces the constraints~\eqref{eq:VoigtKappa} and~\eqref{eq:VoigtMu}, which here result in
\be\label{eq:VoigtBoundsIsotropic}
	\kappa_* \leq f_1\,\kappa_1 + f_2\,\kappa_2, \qquad
	\mu_* \leq f_1\,\mu_1 + f_2\,\mu_2.
\ee
Via the Hashin-Shtrikman variational inequalities from Section~\ref{sec:HS}, we can obtain stronger upper bounds. The variational inequality~\eqref{eq:HSLowerInequality} implies the Hashin-Shtrikman upper bounds
\be\label{eq:HSUpperBoundN}
	\kappa_* \ \leq \
	f_1\kappa_1 + f_2\kappa_2 - \frac{f_1 f_2 (\kappa_1-\kappa_2)^2}
	{f_2\kappa_1+f_1\kappa_2+\frac{4}{3}\mu_+}
\ee
and
\be\label{eq:HSUpperBoundN2}
	\mu_* \ \leq \ f_1\mu_1 + f_2\,\mu_2 - \frac{f_1 f_2(\mu_1 - \mu_2)^2}{f_2\mu_1 + f_1\mu_2 + \mu_+(9\kappa_++8\mu_+)/(6\kappa_++12\mu_+)},
\ee
where we introduced the abbreviations $\mu_+=\max\left\{\mu_1,\mu_2\right\}$ and $\kappa_+=\max\left\{\kappa_1,\kappa_2\right\}$. While these bounds were established decades ago by~\citet{HashinShtrikman1963}, \citet{Hill1963} and \citet{Walpole1966}, we have shown here that they also apply strictly if one (or both) of the two phases violate positive-definiteness of their elastic moduli. Since the last term in each of the two upper bounds is negative (assuming strongly-elliptic elastic moduli for pointwise stability), these bounds are indeed more restrictive than the Voigt bounds~\eqref{eq:VoigtBoundsIsotropic}, as can be expected from Section~\ref{sec:HS}.

Alternatively, we can arrive at tighter bounds for isotropic two-phase composites by considering third-order composite bounds, e.g.\ by taking $\bfepsilon$ as the trial field (see e.g. Section 26.2 of~\citep{Milton2002}), which gives the Beran-Molyneux upper bound on the effective bulk modulus~\citep{BeranMolyneux1966} as simplified by~\citet{Milton1981a} to obtain
\be \label{eq:kappaBeranMolyneux}
	\kappa_* \leq f_1\kappa_1 + f_2\kappa_2 - \frac{f_1 f_2 (\kappa_1-\kappa_2)^2}{f_2\kappa_1+f_1\kappa_2+\frac{4}{3}\langle\mu\rangle_\zeta},
\ee
where
\be
	\langle\mu\rangle_\zeta = \zeta_1\mu_1 + \zeta_2\mu_2, \qquad \zeta_2 = 1-\zeta_1
\ee
and $\zeta_1$ is a parameter depending on three-point statistics and satisfying $0\leq\zeta_1\leq 1$, see Section 6.3 in~\citep{Milton2002} and references therein. Suppose now that
\be\label{eq:assumePositive}
	f_2\kappa_1+f_1\kappa_2+\frac{4}{3}\mu_- \geq 0,
\ee
where $\mu_-=\min\left\{\mu_1,\mu_2\right\}$. In this case the denominator in~\eqref{eq:kappaBeranMolyneux} is positive and consequently
\be\label{eq:HSUpperBound}
	\kappa_* \leq f_1\kappa_1 + f_2\kappa_2 - \frac{f_1 f_2 (\kappa_1-\kappa_2)^2}
	{f_2\kappa_1+f_1\kappa_2+\frac{4}{3}\mu_+},
\ee
which is the Hashin-Shtrikman upper bound. Therefore, if this bound becomes negative and~\eqref{eq:assumePositive} holds, then the composite with traction boundary conditions is necessarily unstable.

If on the other hand
\be\label{eq:assumeNegative}
	f_2\kappa_1+f_1\kappa_2+\frac{4}{3}\mu_- \leq 0,
\ee
then 
\be
	f_1\kappa_1 + f_2\kappa_2 - \frac{f_1 f_2 (\kappa_1-\kappa_2)^2}
	{f_2\kappa_1+f_1\kappa_2+\frac{4}{3}\mu_-}
\ee
(which is the other Hashin-Shtrikman bound) exceeds the Voigt bound, and thus the corresponding sphere assemblage attaining this bound is unstable. The Beran bound does not give any improvement over the Voigt bound when
\be\label{eq:BeranUnstable}
	f_2\kappa_1+f_1\kappa_2+\frac{4}{3}\langle\mu\rangle_\zeta <0.
\ee
However, as the Beran bound includes as part of the trial field subspace a constant field, it must always improve upon the Voigt bound. Therefore, the Beran bound corresponds to maximizing the functional over the subspace of trial fields, not to its minimization. Consequently, we have a contradiction unless the composite is unstable. We conclude that composites are necessarily unstable if $\zeta_1$ is such that~\eqref{eq:BeranUnstable} holds. Excluding those unstable composites where~\eqref{eq:BeranUnstable} holds, we again recover the Hashin-Shtrikman upper bound~\eqref{eq:HSUpperBound}. In summary, \emph{for stable isotropic composites the Hashin-Shtrikman upper bound~\eqref{eq:HSUpperBound} always holds}.


\subsection{Lower bounds for isotropic two-phase composites}
\label{sec:LowerBounds}

As shown in Section~\ref{sec:HomogeneousSolids}, stability requires that the effective moduli (of the homogeneous effective medium) satisfy 
\be\label{eq:NeccComposite}
	\mu_* \geq 0 \qquad\text{and}\qquad \kappa_* \geq -4\mu_*/3.
\ee
for the Dirichlet problem (which is also the necessary condition of pointwise stability) and
\be
	\mu_* \geq 0 \qquad\text{and}\qquad \kappa_* \geq 0
\ee
for Neumann boundary conditions. One can further derive lower bounds on the effective bulk modulus via the translation method starting from the classical variational principle or from the Hashin-Shtrikman variational inequality~\eqref{eq:HSLowerInequality} if there exists an elliptic $\dsC_0$ satisfying~\eqref{eq:Choices}; i.e.\ if there exists a (quasiconvex) reference medium with moduli
\be
	\mu_0 \geq 0, \qquad \kappa_0\geq-4\mu_0/3
\ee
such that
\be
\begin{split}
	\mu_0 & = \min \left\{\mu_1,\mu_2\right\} = \mu_-\\
	-4\mu_0/3 \leq \kappa_0 & \leq \min \left\{\kappa_1,\kappa_2\right\} = \kappa_-,
\end{split}
\ee
which will be the case if
\be\label{eq:HSrestriction}
	\min \left\{\kappa_1,\kappa_2\right\} + \frac{4}{3} \min \left\{\mu_1,\mu_2\right\} \geq 0,
\ee
so that one can take
\be
	\kappa_0 = \min\left\{\kappa_1,\kappa_2\right\}.
\ee
The translated medium with moduli $\dsC_*-\dsC_0$ is thus positive-definite and the Reuss lower bound applies, which implies for the effective bulk modulus that
\be\label{eq:GWM}
	(\kappa_* - \kappa_0)^{-1} \leq f_1 (\kappa_1 - \kappa_0)^{-1} + f_2 (\kappa_2 - \kappa_0)^{-1}.
\ee
If we label the phases such that $\mu_1\geq \mu_2$, we have $\mu_0=\mu_2$ and $\kappa_0 = -4\mu_2/3$, which reduces~\eqref{eq:GWM} to
\be
	(\kappa_* + 4\mu_2/3)^{-1} \leq f_1 (\kappa_1 + 4\mu_2/3)^{-1} + f_2 (\kappa_2 + 4\mu_2/3)^{-1},
\ee
which is exactly the Hill-Hashin-Shtrikman lower bound on the effective bulk modulus $\kappa_*$; i.e.\ presuming~\eqref{eq:NeccComposite} and~\eqref{eq:HSrestriction} hold, we have
\be
	\kappa_* \ \geq \ f_1\kappa_1 + f_2\kappa_2 - \frac{f_1 f_2 (\kappa_1-\kappa_2)^2}
	{f_2\kappa_1+f_1\kappa_2+\frac{4}{3}\mu_-}.
\ee
However, this bound is always negative when the modulus of one phase is negative. Therefore, this bound will only be useful with Dirichlet boundary conditions; with Neumann boundary conditions the best lower bound will be $\kappa_*\geq0$. Analogously, the variational inequality~\eqref{eq:HSLowerInequality} implies the effective shear modulus is bounded by
\be
\begin{split}
	\mu_* \geq f_1\mu_1 + f_2\,\mu_2 - & \frac{f_1 f_2(\mu_1 - \mu_2)^2}{f_2\mu_1 + f_1\mu_2 + \mu_-(9\kappa_-+8\mu_-)/(6\kappa_-+12\mu_-)},
\end{split}
\ee
which applies when~\eqref{eq:NeccComposite} and~\eqref{eq:HSrestriction} are satisfied. Otherwise, the best lower bound will be $\mu_*\geq 0$.

We note that these results can also be interpreted as necessary stability conditions for two-phase isotropic composites. Given the upper bounds~\eqref{eq:HSUpperBoundN} and \eqref{eq:HSUpperBoundN2} and knowing that for a Neumann problem stability requires $\mu_*\geq 0$ and $\kappa_*\geq 0$, we conclude that a composite is necessarily unstable if
\be
	f_1\kappa_1 + f_2\kappa_2 - \frac{f_1 f_2 (\kappa_1-\kappa_2)^2}
	{f_2\kappa_1+f_1\kappa_2+\frac{4}{3}\mu_+} < 0 
	\qquad \text{or} \qquad
	f_1\mu_1 + f_2\,\mu_2 - \frac{f_1 f_2(\mu_1 - \mu_2)^2}{f_2\mu_1 + f_1\mu_2 + \mu_+(9\kappa_++8\mu_+)/(6\kappa_++12\mu_+)} <0.
\ee


\subsection{Further bound considerations for two-phase isotropic composites}

Consider a two-phase isotropic composite and let phase $1$ be the non-positive-definite phase ($\kappa_1<0<\kappa_2$). Let us assume pure traction boundary conditions as a worst-case scenario. Stability requires that $\kappa_*$ be less than the Hashin-Shtrikman-Hill bound~\citep{HashinShtrikman1963,Hill1963}
\be
	0 < \kappa_* < \kappa_2 + \frac{f_1}{\cfrac{1}{\kappa_1-\kappa_2}+\cfrac{f_2}{\kappa_2+\frac{4}{3}\mu_+}},
\ee
i.e.
\be
	\frac{1}{\kappa_1-\kappa_2} + \frac{f_2}{\kappa_2+\frac{4}{3}\mu_+} 
	< - \frac{f_1}{\kappa_2}.
\ee
If this condition is violated, then the composite is necessarily unstable. Since $f_1+f_2=1$, we obtain
\be
	\frac{f_1\,\kappa_1}{\kappa_2(\kappa_1-\kappa_2)}
	+ \frac{f_2\left(\frac{4}{3}\mu_+ +\kappa_1\right)}{\left(\kappa_2+\frac{4}{3}\mu_+\right)(\kappa_1-\kappa_2)} <0.
\ee
Multiplying by $\kappa_2\left(\kappa_2+\frac{4}{3}\mu_+\right)(\kappa_1-\kappa_2)<0$ we arrive at
\be	
	f_2\left(\kappa_1+\frac{4}{3}\mu_+\right)\kappa_2+f_1\left(\kappa_2+\frac{4}{3}\mu_+\right)\kappa_1 > 0,
\ee
i.e.
\be\label{eq:boundEq}
	\kappa_1\kappa_2+\frac{4}{3}\mu_+ \left(f_1\kappa_1+f_2\kappa_2\right) > 0.
\ee
Define $y_\kappa$ by
\be
	\kappa_* = f_1\kappa_1+f_2\kappa_2 - \frac{f_1\,f_2\,(\kappa_1-\kappa_2)^2}{f_2\kappa_1+f_1\kappa_2+y_\kappa}.
\ee
Clearly, for stability of the traction boundary value problem we need $\kappa_*>0$ and $f_2\kappa_1+f_1\kappa_2+y_\kappa>0$. Also~\eqref{eq:boundEq} implies that $f_1\kappa_1+f_2\kappa_2>0$, so that
\be
	\frac{f_1\,f_2\,(\kappa_1-\kappa_2)^2}{f_1\kappa_1+f_2\kappa_2}
	< f_2\kappa_1 + f_1\kappa_2 + y_\kappa
\ee
implying
\be
	\frac{-f_1\,f_2\,(\kappa_1-\kappa_2)^2 + (f_2\kappa_1+f_1\kappa_2) (f_1\kappa_1+f_2\kappa_2)}{f_1\kappa_1 + f_2\kappa_2} + y_\kappa > 0,
\ee
i.e.
\be
	\frac{\kappa_1\,\kappa_2}{f_1\kappa_1 + f_2\kappa_2} + y_\kappa >0.
\ee
In conjunction with the Beran bound~\eqref{eq:kappaBeranMolyneux} which implies $\frac{4}{3}\langle\mu\rangle_\zeta \geq y_k$, this gives us
\be\label{eq:HSNecUnstable}
	\frac{4}{3}\langle\mu\rangle_\zeta > -\frac{\kappa_1\,\kappa_2}{f_1\kappa_1 + f_2\kappa_2}.
\ee
This result in constraints on $\zeta_1=1-\zeta_2$ if a composite is stable. A composite is necessarily unstable if $\zeta_1$ is such that~\eqref{eq:HSNecUnstable} does not hold. This is an improvement over condition~\eqref{eq:BeranUnstable} since
\be
	f_2\kappa_1 + f_1\kappa_2 - \frac{\kappa_1\,\kappa_2}{f_1\kappa_1 + f_2\kappa_2} = \frac{f_1\,f_2\,(\kappa_1-\kappa_2)^2}{f_1\kappa_1 + f_2\kappa_2}>0.
\ee
So
\be
	- \frac{\kappa_1\,\kappa_2}{f_1\kappa_1 + f_2\kappa_2} > f_2\kappa_1+f_1\kappa_2.
\ee
Therefore, no matter what the value of $\zeta_1$, a composite is certainly unstable if
\be
	\frac{4}{3}\mu_+ < - \frac{\kappa_1\,\kappa_2}{f_1\kappa_1 + f_2\kappa_2}
\ee
which is exactly equivalent to~\eqref{eq:boundEq}.


\section{Conclusions}
\label{sec:Conclusions}

We have reviewed the classical variational principles of linear elasticity for heterogeneous solids and composites in the presence of negative-stiffness phases (i.e.\ phases having non-positive-definite elastic moduli), and we have applied those principles to obtain rigorous bounds on the effective elastic moduli. In particular, we have shown the following:
\begin{itemize}
	\item The classical variational principles for Dirichlet and Neumann boundary value problems still apply if the composite is statically stable but the dual variational principles do not apply.
	\item The existence of a unique minimum to the variational principle with Dirichlet boundary conditions is necessary and
sufficient for stability (see Section~\ref{sec:Sufficiency} for a more precise statement of sufficiency).
	\item The classical Hashin-Shtrikman variational inequalities and associated variational principles hold if the composite is statically stable but the associated dual Hashin-Shtrikman principles do not apply.
	\item Stability requires that the Voigt average bounds the effective bulk and shear moduli from above.
	\item Stability further requires that the Hashin-Shtrikman upper bounds for both effective bulk and shear moduli apply.
	\item These bounds imply that, for those cases investigated here, a negative stiffness phase in linear elastic composites or inhomogeneous bodies cannot lead to extreme effective moduli (surpassing those of their constituents) if the composite or body is to be statically stable. This applies to generally anisotropic heterogeneous solids with arbitrary arrangement of the elastic phases and arbitrary variations of their elastic moduli, and it applies equally to Dirichlet and Neumann boundary conditions.
\end{itemize}
While previous investigations derived explicit effective moduli and stability conditions for specific composite geometries or constituent moduli, we have presented here the first investigation resulting in rigorous bounds that limit the space of elastic moduli attainable by composites having negative-stiffness phases.


\subsection*{Acknowledgements}

D.~M.~Kochmann acknowledges support from NSF through CAREER award CMMI-1254424. G.~W.~Milton acknowledges NSF support through grant DMS-1211359.


\vskip 0.5cm
\bibliographystyle{elsarticle-harv}

\end{document}